\documentclass[a4paper,11pt]{article}
\pdfoutput=1 
\usepackage{amssymb,mathrsfs}
\usepackage{bbm,bm,bbding}
\usepackage{siunitx}
\usepackage{amsmath}
\usepackage{braket}
\usepackage{mathtools}
\usepackage[usenames,dvipsnames,svgnames,table]{xcolor}
\usepackage [utf8] {inputenc}
\usepackage{color}
\usepackage{lmodern}
\usepackage{footnote}
\usepackage{slashed}
 \usepackage{mathrsfs}
 \usepackage[pdftex] {graphicx}
\usepackage{multirow}
\usepackage{jheppub} 
\usepackage{here}
\usepackage{hyperref}
\usepackage[justification=centering, singlelinecheck=false]{caption}
\usepackage[list=true, labelfont=bf, labelformat=brace, position=top]{subcaption}
\newcommand{\eq}{\begin{eqnarray}}
\newcommand{\en}{\end{eqnarray}}

\title{Finite-volume effects in the {\boldmath$\delta$}-regime}

\author[1,2,3]{Ulf-G. Mei{\ss}ner,}

\affiliation[1]{Helmholtz-Institut f\"ur Strahlen- und Kernphysik (Theorie)\\ and Bethe Center for Theoretical Physics, Universit\"at Bonn, 53115 Bonn, Germany}                                                                                
\affiliation[2]{Institute for Advanced Simulation (IAS-4),         
 Forschungszentrum J\"ulich,\\ D-52425                     
  J\"ulich, Germany}                                                            
                                                                                
\affiliation[3]{Tbilisi State  University,  0186 Tbilisi, Georgia}              

\emailAdd{meissner@hiskp.uni-bonn.de}

\author[1]{Fabian M\"uller,}

\emailAdd{f.mueller@hiskp.uni-bonn.de}

\author[1,3]{Akaki Rusetsky}
\emailAdd{rusetsky@hiskp.uni-bonn.de}

\abstract{
  We derive a systematic perturbative expansion for the finite-volume energy spectrum
  of the non-linear $O(N)$ $\sigma$-model in the $\delta$-regime. The violation of the
  power-counting rules that emerges after the separation of the fast and slow modes
  is dealt with to all orders by use of the threshold expansion. The known result for the
  rest-frame energy spectrum up-to-and-including next-to-next-to-leading order is
  reproduced.

  \noindent

}

\allowdisplaybreaks
\begin{document}
\maketitle
\flushbottom

\section{Introduction}

In his seminal paper~\cite{Leutwyler:1987ak}, Leutwyler has studied the finite-volume
excitation spectrum of QCD in the chiral limit. It was demonstrated that
a tower of excitations with a zero three-momentum emerges, whose energy in a
box of a size $L$ is proportional to $L^{-3}$. The excitations with such a small energy
cannot be treated in perturbation theory. Rather, they are described by the Lagrangian
of the quantum-mechanical rigid rotator. Besides these ``slow'' modes, the pion field
in the effective Lagrangian of QCD contains the so-called ``fast'' modes (the modes
having nonzero three-momenta), whose energy is proportional to $L^{-1}$. These
modes are perturbative and can be treated by using the conventional diagrammatic
technique.

In the same paper, different regimes of the low-energy expansion of QCD are
identified, depending on the values of the quark masses as well as the external
parameters $L$ (the size of the three-dimensional box) and $T$ (the temperature).
All these expansions are carried out in powers of $1/(FL)^2$, 
where $F$ denotes the pion decay constant in the chiral limit that defines the
hard scale of the theory. 
Assume that the quark mass is nonzero and let $M$ denote the pion
mass at lowest order (for simplicity, we restrict ourselves to the isospin-symmetric world with
two light quarks). If both the temperature $T$ and $M$ are of order of $1/L$, we are in the
so-called $p$-regime, in which the coefficients in the low-energy expansion are functions
of $ML$ and $LT$. For the smaller quark mass, with $M\ll T\sim L^{-1}$ and
$F^2M^2L^3/T\sim 1$, the so-called $\epsilon$-regime sets in, with the low-energy
expansion rearranged. Finally, the $\delta$-regime is characterized by the counting rules
\eq
1/L=O(\delta)\, ,\quad\quad T=O(\delta^3)\, ,\quad\quad M=O(\delta^3)\, .
\en
Here, $\delta$ denotes a generic small parameter.

From the above-mentioned cases, the $p$-regime is the best explored one.
A systematic perturbative approach has been
formulated in Ref.~\cite{Gasser:1987zq} and is routinely used to evaluate finite-volume
artifacts in the observables measured on the lattice.\footnote{The case $ML\gg 1$ and $T\ll M$ can be also attributed to the $p$-regime.
Scattering observables (e.g., two-body scattering phase shifts, weak and electromagnetic decay amplitudes, timelike form factors, etc.) are measured on the lattice in this regime.
The L\"uscher finite-volume
approach~\cite{Luscher:1990ux} has become a standard tool to analyze lattice data
on two-particle scattering and a three-particle framework has been recently proposed
as well~\cite{Hansen:2014eka,Hansen:2015zga,Hammer:2017uqm,Hammer:2017kms,Mai:2017bge}.}
The temperature dependence in this
regime is studied, for example, in Ref.~\cite{Gasser:1986vb}.
Furthermore, a systematic perturbative technique, which has to be used in the
$\epsilon$-regime, has been also set up in Ref.~\cite{Gasser:1986vb}. The subsequent
works~\cite{Hansen:1990yg,Hasenfratz:1989pk} used this technique to study the
temperature and volume dependence of different fundamental characteristics of
the effective field theories (EFT) of QCD at low energy, whereas in
Ref.~\cite{Bedaque:2004dt}
the same approach has been utilized to calculate the nucleon mass shift in a finite volume
and at a finite temperature. The main difference between the calculations in the $p$-
and $\epsilon$-regimes consists in a different treatment of the slow mode. While in
the $p$-regime all modes are treated on equal footing, the global slow mode is singled
out in the $\epsilon$-regime. The partition function in this case is given as a
group average of an expression, in which the slow mode is removed.

The foundations for the perturbative framework in the $\delta$-regime have been laid,
e.g., in Refs.~\cite{Leutwyler:1987ak,Hasenfratz:1993vf,Hasenfratz:2009mp}.
As already mentioned, this regime is relevant for the study of the chiral limit in QCD,
see, e.g., the recent work~\cite{Matzelle:2015lqk}, as
well as an attempt to determine some of the $O(p^4)$ low-energy constants via
the simulations carried out in the $\delta$-regime~\cite{Bietenholz:2010az}.
Perhaps an even more interesting application of the framework
is the study of the volume dependence in different
models of condensed matter physics, which feature massless excitations in the spontaneously broken phase.
For example, the long-wavelength physics of the undoped antiferromagnets is
described by the $O(N)$ non-linear
$\sigma$-model~\cite{Haldane:1983ru,Haldane:1982rj,Chakravarty:1989zz}.
A perturbative framework for these kind of models, which goes under the
name of Magnon Chiral Perturbation
Theory, is formally equivalent to the one used to describe QCD in the chiral limit.   
In the selected papers given below we tried to credit the important work carried out
so far in this
field~\cite{Hasenfratz:1984jk,Neuberger:1988bx,Weingart:2010yv,Niedermayer:2010mx,Niedermayer:2017uyr,Balog:2019vmt,Niedermayer:2018rdw,Hofmann:2002wh,Hofmann:2012ms,Gerber:2009rd,Bar:2003ip}.
Furthermore, it was shown that the hole-doped antiferromagnets with the
spontaneously broken $SU(2)$ spin symmetry in the long-wavelength limit are described
by the Chiral Perturbation Theory of magnons and holes (massive fermions), in a
direct reminiscence of the Baryon Chiral Perturbation
Theory~\cite{Kampfer:2005ba,Brugger:2006dz,Brugger:2009zz,Wiese:2009zz,Chandrasekharan:2006wn,Vlasii:2018rwr}.

Despite the significant effort invested in the study of the $\delta$-regime in the last
decades, the formalism used  still contains caveats which need to be addressed.
Namely, in the $p$- and $\epsilon$-regimes, consistent Feynman rules can be written
down that enable one to carry out calculations (in principle) to any given order in the
expansion. Namely, in the $p$-regime, a single momentum scale $p\sim 1/L$ is relevant
and, hence, the expansion can be carried out without further ado.
Furthermore, the global slow mode in the $\epsilon$-regime can be
straightforwardly taken into account with the use of the Faddeev-Popov trick.
To the contrary, the slow mode in the $\delta$-regime is time-dependent and corresponds
to a dynamical degree of freedom, whose energy scales as $1/L^3$ in the chiral
limit (we remind the reader that the energies of the fast modes scale as $1/L$, see also
a related discussion in Ref.~\cite{Matzelle:2015lqk}). Stated differently, we face a
well-known situation of the EFT with two
distinct scales that need to be clearly separated in order to yield a consistent framework.
In addition, note that the standard Feynman diagrammatical method cannot be used for
the slow mode that makes the separation of scales technically challenging, albeit a
perturbative expansion of the Green functions still can be carried out (we comment in
detail on this rather subtle issue in this paper). In the literature, one finds examples of
calculations of the higher-order corrections~\cite{Leutwyler:1987ak,Hasenfratz:1993vf,Hasenfratz:2009mp,Weingart:2010yv,Niedermayer:2010mx}, 
originating from the loop corrections with the fast modes alone. However, even if, in some cases, the
contribution from the slow mode can be relegated to a rather high order by using a
very special field transformation~\cite{Hasenfratz:1993vf,Hasenfratz:2009mp},
carrying out a systematic expansion to all orders still remains a challenge. The
present work aims at closing this gap.

In this paper we shall demonstrate that a consistent perturbative framework in the
$\delta$-regime emerges by using the so-called ``threshold expansion'' in all diagrams
containing fast as well as slow modes~\cite{Beneke:1997zp}.
In order to set the stage, we shall concentrate on the $O(N)$ $\sigma$-model
and try to reproduce the result of Refs.~\cite{Hasenfratz:1993vf,Hasenfratz:2009mp}
with the use of the alternative technique proposed  here
(the generalization to other models is relatively straightforward and will not be  considered
in this work). It will  in particular be demonstrated that the field redefinition introduced in
Refs.~\cite{Hasenfratz:1993vf,Hasenfratz:2009mp} is in fact unnecessary -- the threshold
expansion neatly does the job (in fact, it is not clear, whether a similar field redefinition
can be used in higher orders as well). We would like to stress that we do not aim to
obtain new results here, pursuing the calculations to even higher orders. Our aim is
rather to set systematic rules that enable carrying out calculations to an arbitrary order
without using additional tricks and which could be eventually extended to the case
where massive fermions are also present.

The paper is organized as follows. In Sect.~\ref{sec:path} we
briefly review the existing approach in case of the $O(N)$ $\sigma$-model and write down
the path integral representation of the Green functions we want to calculate. The low-lying
spectrum at the leading order is considered in the rest-frame as well as moving frames.
In Sect.~\ref{sec:threshold} we show that the results of
Refs.~\cite{Hasenfratz:1993vf,Hasenfratz:2009mp} can be reproduced in any field
parameterization, provided the threshold expansion is used.
Sect.~\ref{sec:concl} contains our conclusions and outlook. Various technical
details are relegated to the appendices.

\pagebreak

\section{Path integral representation of the Green functions, the finite-volume spectrum
  and matrix elements}
\label{sec:path}

\subsection{The Lagrangian and the observables}
\label{sec:observables}

The (Euclidean) Lagrangian of the $O(N)$ $\sigma$-model, which will be used to
demonstrate our method, is given by the following expression
\eq\label{eq:L}
\mathscr{L}=\mathscr{L}^{(2)}+\mathscr{L}^{(4)}+\cdots\, ,
\en
The lowest-order  (LO) Lagrangian is given by:
\eq\label{eq:LO}
\mathscr{L}^{(2)}=\frac{F^2}{2}\,\partial_\mu S^\alpha\partial_\mu S^\alpha\, ,
\en
Here, $S^\alpha,\,\alpha=0,1,\ldots,N-1$, denotes a unit $N$-component real scalar field
that transforms under
$O(N)$. Note also that the above Lagrangian does not contain chiral symmetry
breaking terms and thus describes exactly massless Goldstone bosons. With a slight
abuse of language, we shall call these particles ``pions'' for brevity.

At next-to-leading order (NLO) two additional terms arise that contain the low-energy constants
$\ell_1,\ell_2$:
\eq\label{eq:NLO}
\mathscr{L}^{(4)}=-\ell_1(\partial_\mu S^\alpha\partial_\mu S^\alpha)
(\partial_\nu S^\beta \partial_\nu S^\beta)
-\ell_2(\partial_\mu S^\alpha\partial_\nu S^\alpha)(\partial_\mu S^\beta\partial_\nu S^\beta)\, .
\en
All information about the observables is encoded in the Green functions of the field
$S^\alpha(x)$. For example, the single-particle spectrum in a finite volume
can be extracted from the two-point function in the following manner. Following Ref.~\cite{Luscher:1991wu},
in the rest-frame we consider the correlator\footnote{On the lattice as
explained in Ref.~\cite{Luscher:1991wu}, one should impose  free boundary conditions in the time direction.
This ensures that the states corresponding to the different excited levels of the
rigid rotator do not mix. In our analytic calculations the time elongation of the lattice
is taken to be infinite.}
\eq\label{eq:C_int}
C(t)=\frac{1}{L^6}\,\int d^3\bm{x}\,d^3\bm{y}\,\langle S^\alpha(x)S^\alpha(y)\rangle\, ,\quad\quad t=x_0-y_0\, .
\en
Here, the integration over the spatial dimensions projects onto the states with the zero
total three-momentum and is carried out in a finite cube.
For large Euclidean time separations, one has\footnote{The suppression factor of the remainder in Eq.~(\ref{eq:C1}) will be justified below, see Sect.~\ref{sec:twopoint}.}
\eq\label{eq:C1}
C(t)=\mbox{const}\cdot e^{-M(L)t}+O\left(\exp\left(-\frac{4\pi t}{L}\right)\right)\, .
\en
It is tempting to interpret $M(L)$ as the pion ``mass'' in a box of size $L$.
This interpretation comes, however, with a grain of salt, see below.

Furthermore, in the theory described by the Lagrangian~(\ref{eq:L}), the correlator
can be expanded in regular perturbation series, where the quantity $1/(FL)^2$ plays
the role of the small parameter
\eq\label{eq:C2}
C(t)=1+\sum_{i=0}^\infty\frac{1}{(FL)^{2(i+1)}}\,C_i\left(\frac{t}{L}\right)\, .
\en
Anticipating that $M(L)$ scales like $1/(F^2L^3)$, we expand the first term
in the correlator
in powers of $M(L)t$ (the remainder is still exponentially suppressed, since $t/L\sim 1$
is assumed). Comparing this expansion with Eq.~(\ref{eq:C2}), one may further conclude
that the coefficients of the expansion are polynomials in $t/L$:
\eq
C_i\left(\frac{t}{L}\right)=\sum_{k=0}^{i+1}C_{ik}\left(\frac{t}{L}\right)^k
+O\left(\exp\left(-\frac{4\pi t}{L}\right)\right)\, .
\en
From Eq.~(\ref{eq:C1}) it is also seen that the expansion of $\ln C(t)$ should be linear in
$t$ and the higher-order terms must cancel. The coefficient in front of the linear term
determines $M(L)$, whereas the cancellation of the higher-order terms provides a useful
check in the calculations.

Furthermore, one could try to generalize the above method to  moving frames
with arbitrary three-momentum $\bm{p}$. The counterpart of Eq.~(\ref{eq:C_int}) reads
\eq\label{eq:Cp_int}
C(t,\bm{p})=\frac{1}{L^6}\,\int d^3\bm{x}\,d^3\bm{y}\,
e^{-i\bm{p}(\bm{x}-\bm{y})}\langle S^\alpha(x)S^\alpha(y)\rangle\, .
\en
The behavior of the correlation function at large values of $t$ is given by 
\eq
C(t,\bm{p})=\mbox{const}\cdot e^{-E(L,\bm{p})t}+\cdots\, ,
\en
where the ellipses stand for the exponentially suppressed contributions.

In moving frames, the ground-state energy $E(L,\bm{p})$ scales as $1/L$, and so
does the energy gap between the ground state and the lowest excited level. The logarithm
of $C(t,\bm{p})$ will still be a linear function of $t$ up to exponential corrections.
One could use this property and drop these exponential corrections ``by hand''. In the
remainder, the energy $E(L,\bm{p})$ is still given by the coefficient in the linear
term. Finally, a similar method can be applied for extracting the $k$-pion energy levels
(both in the center-of-mass and moving frames). The simplest way to achieve this
is to use a product of $k$ pion fields
$O^{\alpha_1\cdots\alpha_k}(x)=S^{\alpha_1}(x)\cdots S^{\alpha_k}(x)$.
From $O^{\alpha_1\cdots\alpha_k}(x)$ one could construct a traceless
fully symmetric tensor by subtracting traces with respect to each pair of indices.
This tensor transforms as a basis vector of an
irreducible representation of $O(N)$, which is not contained in the product of
$m$ fundamental representations for any $m<k$. Other observables, like the
magnetization and susceptibility, can also be expressed in terms of the Green functions
and evaluated in perturbation theory in a form of a regular expansion in $1/(FL)^2$.

The main challenge that emerges in the construction of the perturbative series
in $1/(FL)^2$ is the emergence of the so-called slow mode (or, zero mode),
which leads to the ill-defined terms in
standard Feynman diagrams. In order to arrive at the Feynman diagrams without
singularities, the contribution of the slow mode should be treated separately from the
rest. In the following section, we shall briefly discuss how this goal should be achieved.

\subsection{Separation of the slow modes}

The technique for the removal of the slow mode is well known in the literature, see,
e.g.,~\cite{Hasenfratz:1993vf,Hasenfratz:2009mp,Weingart:2010yv}. In order to keep
the presentation self-contained, we shall briefly recapitulate crucial points of the
derivation below. The collective coordinate, corresponding to the time-dependent
net magnetization $m^\alpha(t)$, will be singled out by using the Faddeev-Popov
trick.\footnote{In the $\epsilon$-regime, with $L\sim T^{-1}$, the net magnetization
$m^\alpha$ is a constant. For $T^{-1}\gg L$, however, this approximation is no more valid
and the time-dependent $m^\alpha(t)$ corresponds to a dynamical degree of freedom.}
To this end, we insert the following identity into the path integral
\eq
1=\prod_t\int d^Nm(t)\prod_{\alpha=0}^{N-1}\delta\left(m^\alpha(t)
-\frac{1}{L^3}\,\int d^3\bm{x}\,S^\alpha(\bm{x},t)\right)\, .
\en
Next, one defines the unit vector $ e^\alpha(t)$, and
\eq
m^\alpha(t)=m(t)e^\alpha(t)\, ,\quad\quad d^Nm(t)=m^{N-1}(t)dm(t)d^Ne(t)\,
\delta\left(e^\alpha(t)e^\alpha(t)-1\right)\, .
\en
The partition function can be rewritten as follows
\eq
Z&=&\prod_x\int d^NS(x)\delta\left(S^\alpha(x) S^\alpha(x)-1\right)
\exp\left(-A\left[S\right]\right)
\nonumber\\[2mm]
&=&\prod_x\int d^NS(x)\delta\left(S^\alpha(x)S^\alpha(x)-1\right)
\prod_t\int m^{N-1}(t)dm(t)d^Ne(t)\,\delta\left(e^\alpha(t)e^\alpha(t)-1\right)
\nonumber\\[2mm]
&\times&\delta^N\left(m(t)e^\alpha(t)
-\frac{1}{L^3}\,\int d^3\bm{x}\,S^\alpha(\bm{x},t)\right)
\exp\left(-A\left[S\right]\right)\, ,
\en
where
\eq
A\left[S\right]=\int d^4x\mathscr{L}\left[S(x)\right]
\en
denotes the action functional. We further define the $O(N)$ rotation matrix $\Omega(t)$
as
\eq\label{eq:Omega}
e^\alpha(t)=\Omega^{\alpha\beta}(t)n^\beta\, ,\quad\quad n=(1,0,\ldots,0)\, ,
\en
and choose the following parameterization of $\Omega(t)$
\eq
\Omega^{00}(t)&=&e^0(t), \quad\quad
\Omega^{i0}(t)=e^i(t)\, ,\quad\quad
\Omega^{0i}(t)=-e^i(t)\, ,\quad\quad
\nonumber\\[2mm]
\Omega^{ij}(t)&=&\delta^{ij}-\frac{e^i(t)e^j(t)}{1+e^0(t)}\, ,\quad\quad
i,j=1,\ldots,N-1\, .
\en
Introducing now the new variable $S^\alpha(x)=\Omega^{\alpha\beta}(t)R^\beta(x)$,
the partition function can be rewritten as
\footnote{\label{foot:Sigma}The matrix $\Omega(t)$ is not defined unambiguously, since the vector
$n$ is invariant under $O(N-1)$ group transformations in the subspace orthogonal
to it. This property was used in
Refs.~\cite{Hasenfratz:1993vf,Hasenfratz:2009mp,Weingart:2010yv}, where
the variable transformation in the path integral was written down as
$S^\alpha(x)=\left(\Omega(t)\Sigma^T(t)\right)^{\alpha\beta}R^\beta(x)$.
The block-diagonal matrix
$\Sigma^T(t)$, which leaves the vector $n$ invariant, was chosen in a particular
form that enabled to significantly simplify the structure of the Lagrangian at the order
one is working. In our paper, we do not resort to such a field redefinition and
reproduce the result of the above-mentioned papers by applying the threshold
expansion in all relevant Feynman integrals.} 
\eq
Z&=&\prod_x\int d^NR(x)\delta\left(R^\alpha(x)R^\alpha(x)-1\right)
\prod_t\int m^{N-1}(t)dm(t)d^Ne(t)\,\delta\left(e^\alpha(t)e^\alpha(t)-1\right)
\nonumber\\[2mm]
&\times&\delta^N\left(m(t)n^\alpha
-\frac{1}{L^3}\,\int d^3\bm{x}\,R^\alpha(\bm{x},t)\right)
\exp\left(-A\left[\Omega R\right]\right)\, .
\en
Here, the invariance of the $\delta$-function with respect of the $O(N)$ group transformations has been used. Next, choosing the parameterization
$R=(R^0,\bm{R})\doteq (\sqrt{1-\bm{R}^2},\bm{R})$, and carrying out
the integration over the variables $R^0(x)$ and $m(t)$, we get
\eq\label{eq:representation}
Z&=&\prod_x\int d\bm{R}(x)\,
\prod_t\int d^Ne(t)\,\delta\left(e^\alpha(t)e^\alpha(t)-1\right)\,J
\nonumber\\[2mm]
&\times&\delta^{N-1}\left(-\frac{1}{L^3}\,\int d^3\bm{x}\,\bm{R}(\bm{x},t)\right)
\exp\left(-A\left[\Omega R\right]\right)\, .
\en
Here, the factor $J$ is given by
\eq
J&=&\prod_x\frac{1}{2\sqrt{1-\bm{R}^2(x)}}\times
\prod_t\left(\frac{1}{L^3}\,
\int d^3\bm{x}\,\sqrt{1-\bm{R}^2(x)}\right)^{N-1}
\nonumber\\[2mm]
&=&\mbox{const}\cdot\exp\left(-\frac{1}{2}\,\delta^4(0)\int d^4x
\ln\left[1-\bm{R}^2(x)\right]\right.
\nonumber\\[2mm]
&+&\left.(N-1)\,\delta(0)\int d^4x\ln\left[1-\bm{R}^2(x)\right]\right)\, .
\en
It is immediately seen that the argument of the exponent vanishes in  dimensional
regularization. For this reason, we shall set the factor $J$ equal to one in the following
(the physical results, of course, do not depend on the regularization used).
Note that the path integral representation for a generic Green function looks
similar to Eq.~(\ref{eq:representation}) -- in this case the integrand, in addition, contains the product
of the pertinent operators, expressed in terms of the variables $\bm{R}(x)$ and
$e^\alpha(t)$. The latter correspond to the fast and slow modes, respectively.
The representation displayed in Eq.~(\ref{eq:representation}) provides the basis for
the perturbative expansion of the Green function in powers of $1/L$ (modulo logarithms).
  
  \subsection{Expansion of the Lagrangian}

  In order to formulate the Feynman rules, one has to expand the Lagrangian and
  separate the ``free''  part from the rest. The expansion proceeds along the
  standard pattern and the ``free'' Lagrangian is given by
\eq\label{eq:L-free}
\mathscr{L}_{\sf free}&=&\frac{F^2}{2}\,\dot e^\alpha(t)\dot e^\alpha(t)
+\frac{F^2}{2}\,\partial_\mu\bm{R}(x)\partial_\mu\bm{R}(x)\, .
\en
The above Lagrangian represents a sum of the Lagrangian of the rigid rotator and
the free massless field Lagrangian. In the latter, the mode with zero three-momentum
is absent. All other terms obtained in the expansion of the full Lagrangian in powers of
$\bm{R}$ are considered as a perturbation. Note that a rigid
rotator does not describe non-interacting particles in the standard sense. This was the
reason for putting the word ``free'' in quotation marks.

The two-point function of the free fast fields is given by
\eq\label{eq:propagator-i}
\langle R^i(x)R^j(y)\rangle&=&\frac{\delta^{ij}}{F^2}\,
\int\frac{dk_0}{2\pi}\,
\frac{1}{L^3}\sum_{\bm{k}\neq 0}\frac{e^{ik(x-y)}}{k_0^2+\bm{k}^2}
\doteq\frac{\delta^{ij}}{F^2}\,d(x-y)\, .
  \en
  Both the energy $k_0$ and the three-momentum $\bm{k}$ count as $1/L$. Note also
  that the above sum does not contain the zero three-momentum term $\bm{k}=0$.
  These terms are singled out and are described by the slow mode.

  The slow mode $e^\alpha(t)$ depends on the time only. Its
  three-momentum is exactly zero, and its energy counts as $1/L^3$. Thus, one
  anticipates the breaking of the power counting through the loops which contain both
  fast and slow momenta. Furthermore, the exact two-point function of the slow
  fields can easily be written down. However, it is not very useful since Wick's
  theorem for the slow fields cannot be used. The reason for this is exactly that the free
  Lagrangian for the slow modes coincides with the one of a rigid rotator and
  not of a harmonic oscillator (for more details, see Appendix~\ref{app:Wick}).

  In the following, we plan to carry out the perturbative expansion step by step and
  identify the place where the counting rule breaks down. In order to set the stage,
  we shall focus exclusively on the calculation of the one-particle spectrum.

\subsection{Two-point function at the lowest order}
\label{sec:twopoint}

We start from the two-point function at lowest order which nicely factorizes:
\eq\label{eq:1-4}
C(x-y)&=&\langle S^\alpha(x)S^\alpha(y)\rangle
\nonumber\\[2mm]
&=&\langle \Omega^{\alpha 0}(x^0)\Omega^{\alpha 0}(y^0)\rangle \langle R^0(x) R^0(y)\rangle
\nonumber\\[2mm]
&+&\langle \Omega^{\alpha i}(x^0)\Omega^{\alpha 0}(y^0)\rangle \langle R^i(x) R^0(y)\rangle
\nonumber\\[2mm]
&+&\langle \Omega^{\alpha 0}(x^0)\Omega^{\alpha i}(y^0)\rangle \langle R^0(x) R^i(y)\rangle
\nonumber\\[2mm]
&+&\langle \Omega^{\alpha i}(x^0)\Omega^{\alpha j}(y^0)\rangle \langle R^i(x) R^j(y)\rangle\, .
\en
In this expression, the vacuum expectation values of the slow and fast variables
are defined as follows
\eq
&&\langle \Omega^{\alpha\beta}(x^0)\Omega^{\gamma\delta}(y^0)\cdots\rangle
=\prod_t\int d^N e(t)\delta(e^\lambda(t)e^\lambda(t)-1) \left(\Omega^{\alpha\beta}(x^0)\Omega^{\gamma\delta}(y^0)\cdots\right)
\nonumber\\[2mm]
&&\quad\quad\quad\times\,\exp\left(-\frac{F^2L^3}{2}\int dt\,\dot e^\sigma(t)\dot e^\sigma(t)\right)\, ,
\nonumber\\[2mm]
&&\langle R^\alpha(x) R^\beta(y)\cdots\rangle
=\prod_x\int d\bm{R}(x)\delta^{N-1}\left(-\frac{1}{L^3}\,
  \int d^3\bm{x}\,\bm{R}(\bm{x},t)\right)
\nonumber\\[2mm]
&&\quad\quad\quad\times\,\left(R^\alpha(x) R^\beta(y)\cdots\right)
\exp\left(-\frac{F^2}{2}\,\int d^4x\partial_\mu\bm{R}(x)\partial_\mu\bm{R}(x)\right)\, .
\en
The second and the third terms in Eq.~(\ref{eq:1-4}) vanish identically, because the
free Lagrangian has the symmetry with respect to $\bm{R}\to -\bm{R}$. 
The fast mode propagator in the fourth term is given in Eq.~(\ref{eq:propagator-i}),
whereas the correlator of two composite fields $R^0$
in the first term can be expanded as
\eq
&&\langle R^0(x) R^0(y)\rangle\doteq T_0(x-y)=
\nonumber\\[2mm]
&=&1-\langle\bm{R}^2(0)\rangle
-\frac{1}{4}\,\langle\left(\bm{R}^2(0)\right)^2\rangle
+\frac{1}{4}\,\langle\bm{R}^2(x)\bm{R}^2(y)\rangle
+O(\bm{R}^6)
\nonumber\\[2mm]
&=&1-\frac{N-1}{F^2}\,d(0)-\frac{N-1}{2F^4}\,d^2(0)+\frac{N-1}{2F^4}\,d^2(x-y)
+O(F^{-6})
\nonumber\\[2mm]
&\doteq&\sum_{a=0}^\infty C_a d^{2a}(x-y)\, .
\en
Performing the Fourier transform, one gets
\eq\label{eq:momentum-sums}
C(t,\bm{p})&\doteq&\int d^3\bm{x}e^{-i\bm{p}\bm{x}}C(x)
\nonumber\\[2mm]
&=&\langle \Omega^{\alpha i}(t)\Omega^{\alpha i}(0)\rangle
\frac{1-\delta_{\bm{p}0}}{F^2}\,
\frac{e^{-|\bm{p}|t}}{2|\bm{p}|}
\nonumber\\[2mm]
&+&\langle \Omega^{\alpha 0}(t)\Omega^{\alpha 0}(0)\rangle
\int d^3\bm{x}e^{-i\bm{p}\bm{x}}T_0(\bm{x},t)\, .
\en
In this expression,
\eq\label{eq:T0}
&&\int d^3\bm{x}e^{-i\bm{p}\bm{x}}T_0(\bm{x},t)
=L^3\delta_{\bm{p}0}+\sum_{a=1}^\infty
\frac{C_a}{L^{6a}}\,\sum_{\bm{k}_1\cdots \bm{k}_{2a}}
L^3\delta_{\bm{p},\bm{k}_1+\cdots+\bm{k}_{2a}}
\nonumber\\[2mm]
&&\quad\quad\times\,(1-\delta_{\bm{k}_10})\cdots(1-\delta_{\bm{k}_{2a}0})
\frac{\exp(-(|\bm{k}_1|+\cdots +|\bm{k}_{2a}|)t)}
{2|\bm{k}_1|\cdots2|\bm{k}_{2a}|}\, .
\en
We consider two cases separately:

\begin{itemize}
\item $\bm {p}=0$: In this case, the first term in Eq.~(\ref{eq:momentum-sums})
  does not contribute. Furthermore, in the momentum sums contained in the quantity
  $T_0$, see Eq.~(\ref{eq:T0}), one always has $|\bm{k}|_1+\cdots +|\bm{k}|_{2a}\geq 2$
  (in units of $2\pi/L$). Hence, the leading contribution comes from the constant term,
  and we get:
  \eq
  C(t,\bm{0})=L^3\langle\Omega^{\alpha 0}(t)\Omega^{\alpha 0}(0)\rangle
  +O\left(\exp\left(-\frac{4\pi}{L}\right)\right)\, .
  \en
  Note that the argument of the exponentially suppressed term was already anticipated
  in Eq.~(\ref{eq:C1}).
  
\item $\bm{p}\neq 0$. In this case, the first term in Eq.~(\ref{eq:momentum-sums})
  contributes. Furthermore, $|\bm{k}|_1+\cdots + |\bm{k}|_{2a}\geq |\bm{p}|$ always holds
  for $\bm{k}_1+\cdots+\bm{k}_{2a}=\bm{p}$ and $\bm{k}_i\neq 0$. Note that for
  $|\bm{p}|\geq 2$ the configurations of $\bm{k}_i$ exist that obey the equality
  $|\bm{k}|_1+\cdots +|\bm{k}|_{2a}=|\bm{p}|$.
  We denote the sum over such
  configurations by ${\sum}^\prime$. For all other configurations,
  $|\bm{k}|_1+\cdots +|\bm{k}|_{2a}\geq |\bm{p}|+a$, where $a$ depends on the
  choice of the vector $\bm{p}$.
  Hence, we have:
  \eq
   &&\quad\quad C(t,\bm{p})
=\langle \Omega^{\alpha i}(t)\Omega^{\alpha i}(0)\rangle
\frac{e^{-|\bm{p}|t}}{2F^2|\bm{p}|}
+\langle \Omega^{\alpha 0}(t)\Omega^{\alpha 0}(0)\rangle e^{-|\bm{p}|t}
\nonumber\\[2mm]
&\times&\biggl(\sum_a\frac{C_a}{L^{6a}}\,{\sum\limits_{\bm{k}_1\cdots \bm{k}_{2a}}}^{\hspace*{-.2cm}\prime}\,\,
\frac{L^3\delta_{\bm{p},\bm{k}_1+\cdots+\bm{k}_{2a}}(1-\delta_{\bm{k}_10})\cdots(1-\delta_{\bm{k}_{2a}0})}{2|\bm{k}_1|\cdots2|\bm{k}_{2a}|}
+O(e^{-at})\biggr)\, .\quad\quad
\en
  
\end{itemize}
Furthermore, in order to evaluate the vacuum expectation value of the product of two
$\Omega$'s, we insert a full set of the rigid rotator eigenstates between two operators
\eq
\langle \Omega^{\alpha i}(t)\Omega^{\alpha i}(0)\rangle
&=&\sum_ne^{-\varepsilon_n t}\langle 0| \Omega^{\alpha i}(0)|n\rangle
\langle n|\Omega^{\alpha i}(0)|0\rangle\, ,
\nonumber\\[2mm]
\langle \Omega^{\alpha 0}(t)\Omega^{\alpha 0}(0)\rangle
&=&\sum_ne^{-\varepsilon_n t}\langle 0| \Omega^{\alpha 0}(0)|n\rangle
\langle n|\Omega^{\alpha 0}(0)|0\rangle\, .
\en
Here,
\eq\label{eq:varepsilon_n}
\varepsilon_n=\frac{n(n+N-2)}{2\Theta}\,,\quad\quad n=0,1,\ldots
\en
denotes the $n^{\rm th}$ eigenvalue of the Hamiltonian, corresponding to the
eigenvector $|n\rangle$ and $\Theta=F^2L^3$ is the moment of inertia.
As seen, the eigenvalues scale as $L^{-3}$.

In the next step, we consider the sums over the eigenvectors of the rigid rotator (more details can be
found in Appendix~\ref{app:rotator}). Since $\Omega^{\alpha 0}=e^\alpha$ is the irreducible tensor operator belonging to the fundamental representation of $O(N)$,
all matrix elements $\langle 0|\Omega^{\alpha 0}(0)|n\rangle$ vanish except for $n=1$.
Furthermore, using a proper choice of the basis vectors
in the fundamental representation, $|1,\gamma\rangle\, ,~\gamma=1,\ldots,N$,
the above matrix element takes a simple form
$\langle 0|\Omega^{\alpha 0}(0)|1\gamma\rangle=\delta^{\alpha\gamma}$ (the normalization of the matrix follows from the condition $e^\alpha e^\alpha=1$). The calculation
of the matrix element of $\Omega^{\alpha i}$ is a more complicated task and is
considered in Appendix~\ref{app:rotator}). Here, it suffices to say that the state
with $n=0$ also contributes to the sum over the intermediate states since the
corresponding matrix element does not vanish.

To summarize, the free finite-volume spectrum of the two-point function has rather
peculiar properties. Namely, in the rest-frame, the lowest excitation has the energy
$\varepsilon_1(L)=(N-1)/(2F^2L^3)$ and the first excited level is separated by $4\pi/L$.
In the frame moving with a momentum $\bm{p}$, the lowest energy level is exactly
at $E=|\bm{p}|$. This, by the way, demonstrates that interpreting the
quantity $(N-1)/(2F^2L^3)$ as the pion mass in a finite volume is a slight
abuse of language, because the lowest level in different frames does not obey the
relativistic dispersion law. This does not come at a complete surprise, since the
separation of the fast and slow modes is done in the rest-frame and breaks
Lorentz invariance from the beginning.
  In case of massive particles, there exists a natural scale: if $ML\gg 1$, the fast
  and slow modes glue together and form a relativistic particle. There exists no such
  scale in the massless case and the lattice breaks relativistic invariance at all scales.

\subsection{Perturbative expansion}

At the next step, we evaluate radiative corrections to the two-point function, in order
to obtain a regular expansion of the low-lying spectrum in powers of $1/(FL)^2$.
At NLO, it suffices to work with the Lagrangian
$\mathscr{L}^{(2)}$. Expanding this Lagrangian in powers of the field $\bm{R}$,
one gets:
 \eq\label{eq:L2}
 \mathscr{L}^{(2)}=\mathscr{L}_{\sf free}
 -\frac{F^2}{2}\,\dot e^\alpha\dot e^\alpha\bm{R}^2
+\frac{F^2}{2}\,\dot\Omega^{\alpha i}\dot\Omega^{\alpha j}R^iR^j
+\frac{F^2}{2}\,\left(\dot\Omega^{\alpha i}\Omega^{\alpha j}-\dot\Omega^{\alpha j}\Omega^{\alpha i}\right)R^i\dot R^j+\cdots~.
\nonumber\\
\en
  At this place, it is appropriate to discuss the power counting. The dimensionless field
$e^\alpha(t)$ counts as $O(1)$. From Eq.~(\ref{eq:propagator-i}) one concludes
that the field $\bm{R}(x)$ should count as $O(L^{-1})$ (we remind the reader that all
components of the four-momentum of the fast mode count as $O(L^{-1})$). On the
contrary, the energy of the soft mode counts as $O(L^{-3})$, and its three-momentum is
zero by definition. Consequently, $\dot e^\alpha(t)$ in the Lagrangian counts as $O(L^{-3})$
and $\partial_\mu\bm{R}(x)$ as $O(L^{-2})$. The free Lagrangian of the slow modes,
which will be used to read off the spectrum, is given in
Eq.~(\ref{eq:L-free}) and counts as $O(L^{-6})$. This effective Lagrangian is obtained
by integrating out the fast modes. To this end, one first writes:
\eq\label{eq:exp-L}
\exp\left(-\int d^4x\,\mathscr{L}^{(2)}(x)\right)&=&
1-\frac{1}{1!}\,\int d^4x\,\mathscr{L}^{(2)}(x)
\nonumber\\[2mm]
&+&\frac{1}{2!}\,
\int d^4x\,\mathscr{L}^{(2)}(x)\int d^4y\,\mathscr{L}^{(2)}(y)+\cdots\, ,
\en
integrates out the fields $\bm{R}(x)$ in the path integral and finally exponentiates the
remainder back. Dealing with the first two terms is easy: in the first-order term of Eq.~(\ref{eq:exp-L}) we have
two fields $R^i(x)R^j(x)$ which have to be contracted. This is equivalent
to replacing $R^i(x)R^j(x)$ by $\langle R^i(x)R^j(x)\rangle$ in
the Lagrangian. Graphically, this corresponds to the tadpole diagram shown in
Fig.~\ref{fig:1loop}a. It is immediately seen that the resulting term in the
Lagrangian should count as $O(L^{-8})$ and thus results into a $O(L^{-2})$ correction
to the leading-order result for the energy spectrum. Higher order terms in
perturbation theory (\ref{eq:exp-L}) with this part of the Lagrangian will be more
suppressed and are not considered here. With the last term in Eq.~(\ref{eq:L2}), the
situation is slightly different. The tadpole term does not contribute, because
$\langle\dot R^i(x)R^j(x)\rangle=0$. Integrating out the
fields $\bm{R}(x)$ in the second-order term of Eq.~(\ref{eq:exp-L}) leads to a loop shown in
Fig.~\ref{fig:1loop}b. In addition of the product of four fields $\Omega$ with two time
derivatives that count as $O(L^{-6})$, there are two additional lines of field $\bm{R}(x)$,
two derivatives on these fields and an additional integration over $d^4x$. All this
results in a factor $O(L^{-2})$ in addition, so the contribution of the diagram
Fig.~\ref{fig:1loop}b in the effective Lagrangian counts as $O(L^{-8})$, the same as
the contribution of Fig.~\ref{fig:1loop}a. We would also stress here that the whole
discussion above is based on the naive power counting which gives the leading
power correctly.

In Refs.~\cite{Hasenfratz:1993vf,Hasenfratz:2009mp,Weingart:2010yv},
a field transformation in the Lagrangian given by Eq.~(\ref{eq:L2})
is performed that renders its structure
simpler. We discuss this transformation in detail in Appendix~\ref{app:trafo}.
It has the following structure
\eq\label{eq:Omega-Sigma}
\Omega(t)\to\Omega(t)\Sigma^T(t)\, ,
\en
where
\eq\label{eq:Sigma}
\Sigma^{00}(t)=1\, ,\quad\quad\Sigma^{0j}(t)=\Sigma^{i0}(t)=0\, ,\quad\quad
\Sigma^{ij}(t)=\hat\Sigma^{ij}(t)\, ,
\en
and $\hat\Sigma(t)\in O(N-1)$. Clearly, this transformation leaves the vector
$n=(1,0,\ldots,0)$ invariant, see footnote~\ref{foot:Sigma}. 

It can be shown (see Appendix~\ref{app:trafo}) that, after the field transformation with a
properly chosen matrix $\Sigma$, the last term in Eq.~(\ref{eq:L2}) vanishes:
  \eq
\frac{d}{dt}\,(\Omega(t)\Sigma^T(t))^{\alpha i}(\Omega(t)\Sigma^T(t))^{\alpha j}=0\, .
\en
Furthermore, it can be shown that, after the same transformation,
\eq\label{eq:Hasenfratz}
\frac{d}{dt}\,(\Omega\Sigma^T)^{\alpha i}(t)
\frac{d}{dt}\,(\Omega\Sigma^T)^{\alpha i}(t)=\frac{1}{2}\,\dot e^\alpha(t)\dot e^\alpha(t)\, .
\en
This allows one to use a shortcut in the calculation of the excitation spectrum at NLO. At
this order, only the tadpole diagrams shown in Fig.~\ref{fig:1loop}a contribute, and
the net effect reduces to a correction to the moment of inertia
\eq\label{eq:Theta-NLO}
\Theta\to\Theta\biggl(1-\frac{N-2}{F^2}\,d(0)\biggr)\, .
\en
The excitation spectrum in the rest-frame is given by~\cite{Hasenfratz:2009mp}
\eq
\varepsilon_n=\frac{n(n+N-2)}{2F^2L^3}\,\biggl(1+\frac{N-2}{F^2}\,d(0)\biggr)\,,~n=0,1,\ldots
\en
Note that, according to Eq.~(\ref{eq:propagator-i}), the quantity $d(0)$ is of order
$1/L^2$. No multi-scale integrals arise, and the power counting is not violated. Furthermore, as shown in Ref.~\cite{Hasenfratz:2009mp}, no new structures arise at
NNLO as well, and hence the same method can be used for calculations also in this case.\footnote{It should be mentioned that in Ref.~\cite{Hasenfratz:2009mp} an additional approximation was used, assuming that the terms in the perturbative expansion, which are non-local in time, are exponentially suppressed. In fact, this approximation is conceptually related to the threshold expansion used in the present paper. As we shall show, however,
the suppression is only power-like and will show up in subsequent orders.}
It, however, remains unclear how this field transformation can be systematically
performed at higher orders. On the other hand, physical results cannot depend
on the choice of the interpolating field. This is not immediately manifest in the present
example. If the field transformation is not performed, the loop diagram  (the self-energy)
shown in Fig.~\ref{fig:1loop}b also contributes. 
The integrand of this diagram depends on different low-energy scales, and the power counting is not straightforward. In the following section we shall
demonstrate how the above problems can be addressed.

\section{The threshold expansion}
\label{sec:threshold}

\subsection{NLO}

As became clear from the previous discussion, the main problem that precludes carrying
out a systematic expansion in inverse powers of $L$ is related to the presence of
different scales in the Feynman diagrams. A standard method to address this problem,
applicable at any order,
is to use threshold expansion in these diagrams. This method enables one to arrive at the final
result without further ado (e.g., field redefinitions). The
independence of the result on the choice of the field will then represent a nice test
of the calculations.

\begin{figure}[t]
\begin{center}
  \includegraphics[width=0.55\paperwidth]{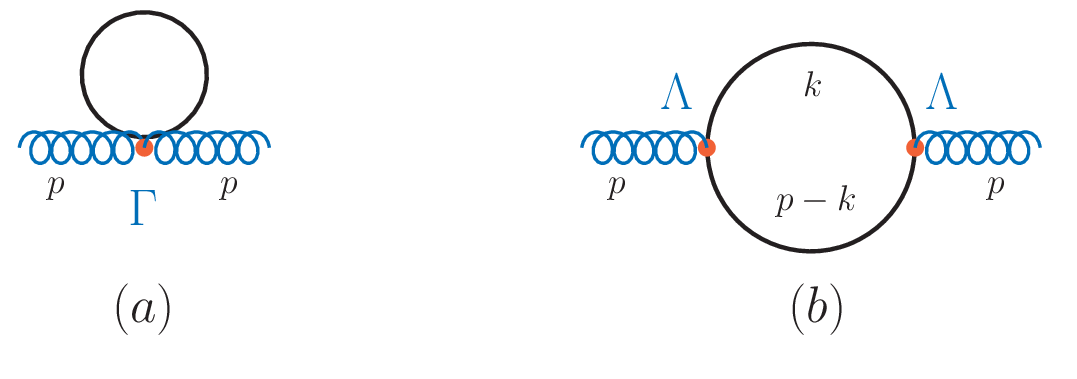}
  \caption{One-loop contributions to the self-energy: a) the tadpole diagram, b) the
    second-order (self-energy)  diagram. The solid and wiggle lines denote the fast and the slow
    modes, respectively. The vertices $\Gamma$ and $\Lambda$ emerge from the Lagrangian (\ref{eq:L2}) after integrating out the fast mode perturbatively, see Eqs.~(\ref{eq:Lambda})
    and (\ref{eq:Gamma}).}
  \label{fig:1loop}
\end{center}  
\end{figure}

We shall first explain this method in detail at NLO. Since we focus on the calculation of the
excitation spectrum in the rest frame, we shall use the same shortcut as in Refs.~\cite{Hasenfratz:1993vf,Hasenfratz:2009mp,Weingart:2010yv}, evaluating the (non-local) effective
action at  second order. The piece of this action that contains the contribution
of the diagram in Fig.~\ref{fig:1loop}b (the culprit) to the pion self-energy, is given by
\eq
\delta S_{\sf eff}^{(b)}=\int d^4x\, \delta\mathscr{L}_{\sf eff}^{(b)}(x)\, ,
\en
where
\eq\label{eq:dL}
\delta\mathscr{L}_{\sf eff}^{(b)}(x)&=&-\frac{F^4}{8}\,\int d^4z\Lambda^{ij}(u_0)\Lambda^{kn}(v_0)
\biggl(\langle R^i(u)R^k(v)\rangle\langle \dot R^j(u)\dot R^n(v)\rangle
\nonumber\\[2mm]
&+&\langle R^i(u)\dot R^n(v)\rangle\langle \dot R^j(u) R^k(v)\rangle\biggr)\, .
\en
Here, $u=x+z/2$, $v=x-z/2$ and
\eq\label{eq:Lambda}
\Lambda^{ij}(u_0)=\dot\Omega^{\alpha i}(u_0)\Omega^{\alpha j}(u_0)
-\dot\Omega^{\alpha j}(u_0)\Omega^{\alpha i}(u_0)\, .
\en

\noindent
We have in addition used the fact that
\eq
\langle R^i(u)\dot R^j(u)\rangle
=\langle R^k(v)\dot R^n(v)\rangle=0\, .
\en
Next, note that
\eq\label{eq:I1}
&&\langle R^i(u)R^k(v)\rangle\langle \dot R^j(u)\dot R^n(v)\rangle
=\frac{\delta^{ik}}{F^2}\,\frac{\delta^{nj}}{F^2}\,I_1\, ,
\nonumber\\[2mm]
I_1&=&\int \frac{dk_0}{2\pi}\,\frac{1}{L^3}\,\sum_{\bm{k}\neq 0}
\int \frac{dp_0}{2\pi}\,\frac{1}{L^3}\,\sum_{\bm{p}\neq\bm{k}}
\frac{k_0^2e^{-ipz}}{k^2(p-k)^2}\, ,
\en
and
\eq\label{eq:I2}
&&\langle R^i(u)\dot R^n(v)\rangle\langle \dot R^j(u) R^k(v)\rangle
=\frac{\delta^{in}}{F^2}\,\frac{\delta^{kj}}{F^2}\, I_2\, ,
\nonumber\\[2mm]
I_2&=&\int \frac{dk_0}{2\pi}\,\frac{1}{L^3}\,\sum_{\bm{k}\neq 0}
\int \frac{dp_0}{2\pi}\,\frac{1}{L^3}\,\sum_{\bm{p}\neq\bm{k}}
\frac{k_0(p_0-k_0)e^{-ipz}}{k^2(p-k)^2}\, .
\en
The effective Lagrangian
$\delta\mathscr{L}_{\sf eff}^{(b)}$ is essentially
non-local. Consider now the loop integrals in Eqs.~(\ref{eq:I1}) and (\ref{eq:I2}).
First, since the matrices $\Lambda^{ij},\Lambda^{ik}$ in Eq.~(\ref{eq:dL}) do not
depend on the argument $\bm{z}$, the integration over this variable can be carried
out and one gets: 
\eq
\int d^3\bm{z}\,I_1&=&\int \frac{dk_0}{2\pi}\,\frac{1}{L^3}\,\sum_{\bm{k}\neq 0}
\int \frac{dp_0}{2\pi}\,\frac{1}{L^3}\,\sum_{\bm{p}\neq \bm{k}}L^3\delta_{\bm{p}0}
\frac{k_0^2e^{-ip_0z_0}}{k^2(p-k)^2}
\nonumber\\[2mm]
&=&\int \frac{dp_0}{2\pi}\,e^{-ip_0z_0}
\int \frac{dk_0}{2\pi}\,\frac{1}{L^3}\,\sum_{\bm{k}\neq 0}
\frac{k_0^2}{k^2((p_0-k_0)^2+\bm{k}^2)}\, .
\en
It is now seen that the above integral features different momentum scales. Namely, the
energy of the slow mode, $p_0$, scales like $L^{-3}$, whereas the three-momentum
of the fast mode, $\bm{k}$, scales like $L^{-1}$. It is easy to ensure that the only
non-vanishing contribution to this integral comes from the region where $k_0$ also
scales like $L^{-1}$. Applying the so-called threshold expansion~\cite{Beneke:1997zp}
to this integral, one finally gets
\eq\label{eq:I1n}
\int d^3\bm{z}\,I_1&=&\int \frac{dp_0}{2\pi}\,e^{-ip_0z_0}
\int \frac{dk_0}{2\pi}\,\frac{1}{L^3}\,\sum_{\bm{k}\neq 0}
\frac{k_0^2}{k^2(k_0^2+\bm{k}^2)}\biggl(1+\frac{2p_0k_0}{k_0^2+\bm{k}^2}+\cdots\biggr)
\nonumber\\[2mm]
&=&\delta(z_0)\int \frac{dk_0}{2\pi}\,\frac{1}{L^3}\,\sum_{\bm{k}\neq 0}
\frac{k_0^2}{(k^2)^2}+\cdots
\nonumber\\[2mm]
&=&\frac{1}{4}\,\delta(z_0)\frac{1}{L^3}\,\sum_{\bm{k}\neq 0}\frac{1}{|\bm{k}|}+\cdots
=\frac{1}{2}\,\delta(z_0)d(0)+\cdots
\en
The integral $I_2$ can be expanded similarly
\eq\label{eq:I2n}
\int d^3\bm{z}\,I_2=-\frac{1}{2}\,\delta(z_0)d(0)+\cdots
\en
Substituting this result in Eq.~(\ref{eq:dL}), one obtains
\eq
&&\delta\mathscr{L}_{\sf eff}^{(b)}(x)=-\frac{1}{16}\,d(0)\Lambda^{ij}(x_0)\Lambda^{kn}(x_0)
(\delta^{ik}\delta^{jn}-\delta^{in}\delta^{jk})+\cdots
\nonumber\\[2mm]
&=&-\frac{1}{4}\,d(0)(\dot\Omega^{\alpha i}(x_0)\Omega^{\alpha j}(x_0)
-\dot\Omega^{\alpha j}(x_0)\Omega^{\alpha i}(x_0))
\dot\Omega^{\beta i}(x_0)\Omega^{\beta j}(x_0)+\cdots
\nonumber\\[2mm]
&=&-\frac{1}{2}\,d(0)\dot\Omega^{\alpha i}(x_0)\Omega^{\alpha j}(x_0)
\dot\Omega^{\beta i}(x_0)\Omega^{\beta j}(x_0)+\cdots
\nonumber\\[2mm]
&=&-\frac{1}{2}\,d(0)(\dot\Omega^{\alpha i}(x_0)\dot\Omega^{\alpha i}(x_0)
-\dot\Omega^{\alpha 0}(x_0)\dot\Omega^{\alpha 0}(x_0))+\cdots
\en
This expression should be added to the tadpole contribution coming from the diagram
in Fig.~\ref{fig:1loop}a. The corresponding effective Lagrangian is local and is given by
\eq
\delta\mathscr{L}_{\sf eff}^{(a)}(x)&=&
\frac{F^2}{2}\,\Gamma^{ij}(x_0)\langle R^i(x)R^j(x)\rangle
\nonumber\\[2mm]
&=&-\frac{N-1}{2}\,d(0)\dot e^\alpha(x_0)\dot e^\alpha(x_0)
+\frac{1}{2}\,d(0)\dot\Omega^{\alpha i}(x_0)\dot\Omega^{\alpha i}(x_0)\, ,
\en
where
\eq\label{eq:Gamma}
\Gamma^{ij}(x_0)=-\dot e^\alpha(x_0)\dot e^\alpha(x_0)
+\dot\Omega^{\alpha i}(x_0)\dot\Omega^{\alpha j}(x_0)\, .
\en
Hence,
\eq\label{eq:ab}
\delta\mathscr{L}_{\sf eff}^{(a)}(x)+\delta\mathscr{L}_{\sf eff}^{(b)}(x)
=-\frac{N-2}{2}\,d(0)\dot e^\alpha(x_0)\dot e^\alpha(x_0)\, ,
\en
and the result given in Eq.~(\ref{eq:Theta-NLO}) is readily reproduced.
Hence, as expected, the physical result
does not depend on the field parameterization.\footnote{Note that the final result
contains only $\Omega^{\alpha 0}(t)$ and thus does not depend on the
matrix $\Sigma$ in Eq.~(\ref{eq:Omega-Sigma}).} In other words, the field
transformation, introduced in
Refs.~\cite{Hasenfratz:1993vf,Hasenfratz:2009mp,Weingart:2010yv}, ensures the
separation of the fast and slow modes at the order one is working.

\subsection{NNLO}

The interaction Lagrangian that will be used in the calculations at NNLO is obtained
by the expansion of the Lagrangian given in Eqs.~(\ref{eq:L})-(\ref{eq:NLO}) in powers
of the field $\bm{R}$. Up to the NNLO, the relevant terms are given by
  \eq\label{eq:L4}
  \mathscr{L}^{(4)}&=&\frac{F^2}{2}\,\Gamma^{ij}R^iR^j+\frac{F^2}{2}\,\Lambda^{ij}R^i\dot R^j
  -F^3\Delta^i\bm{R}^2\dot R^i+\mathscr{L}_R
\nonumber\\[2mm]
  &-&4\ell_1 b_1^{ij}\dot R^i\dot R^j
  -2\ell_2 b_2^{ij}\dot R^i\dot R^j+\cdots\, ,\quad
\en  
where
\eq
\Gamma^{ij}&=&-\delta^{ij}\dot e^\alpha\dot e^\alpha+\dot\Omega^{\alpha i}\dot\Omega^{\alpha j}\, ,
\nonumber\\[2mm]
\Lambda^{ij}&=&\dot\Omega^{\alpha i}\Omega^{\alpha j}-\dot\Omega^{\alpha j}\Omega^{\alpha i}\, ,
\nonumber\\[2mm]
\Delta^i&=&\dot\Omega^{\alpha 0}\Omega^{\alpha i}-\dot\Omega^{\alpha i}\Omega^{\alpha 0}\, ,
\nonumber\\[2mm]
b_1^{ij}&=&\dot\Omega^{\alpha 0}\Omega^{\alpha i}\dot\Omega^{\beta 0}\Omega^{\beta j}\, ,
\nonumber\\[2mm]
b_2^{ij}&=&\dot\Omega^{\alpha 0}\dot\Omega^{\alpha 0}\Omega^{\beta i}\Omega^{\beta j}+\dot\Omega^{\alpha 0}\dot\Omega^{\beta 0}\Omega^{\alpha i}\Omega^{\beta j}\, ,
\nonumber\\[2mm]
\mathscr{L}_R&=&\left(\bm{R}\partial_\mu \bm{R}\right)^2\, .
\en
Note that $\Gamma^{ij}$ and $\Lambda^{ij}$ have been introduced earlier, in Eqs.~(\ref{eq:Gamma}) and (\ref{eq:Lambda}), respectively.
In the calculations at the NLO, only the first two terms of Eq.~(\ref{eq:L4}) contribute.

\begin{figure}[t]
  \begin{center}
    \includegraphics*[width=12.cm]{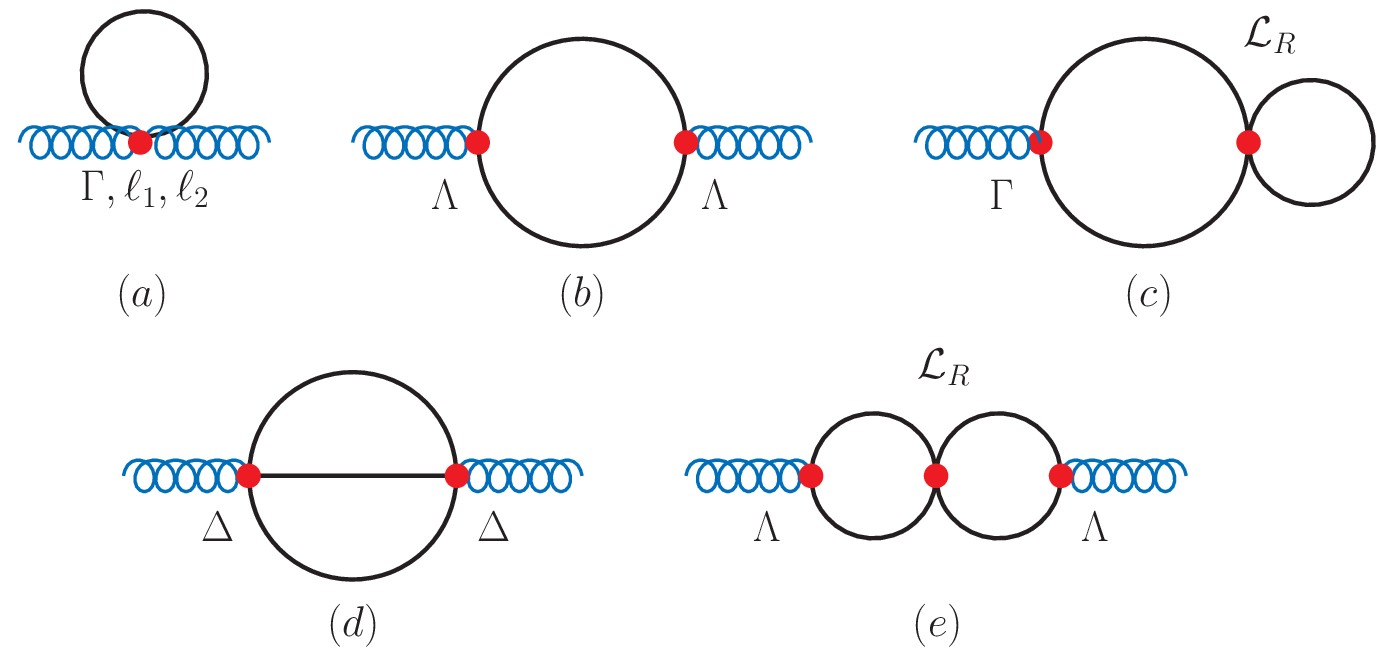}
    \caption{The diagrams contributing to the self-energy at NNLO. For notations, see Fig.~\ref{fig:1loop}.}
    \label{fig:NNLO}
  \end{center}
  \end{figure}

Since we shall again focus on the calculation of the excitation spectrum in the
rest-frame, the same shortcut as at the NLO can be used as a substitute for the general method
described in Sect.~\ref{sec:observables}. Only the diagrams depicted in Fig.~\ref{fig:NNLO} contribute.
The calculations based on the threshold expansion are pretty standard
and will not be described in much detail.

\subsubsection*{ Figure~\ref{fig:NNLO}a, $\Gamma$ + Figure~\ref{fig:NNLO}b, $\Lambda\Lambda$}
The sum of two diagrams shown in Fig.~\ref{fig:NNLO}a (with the vertex $\Gamma$)
and in Fig.~\ref{fig:NNLO}b has been calculated already, see Eq.~(\ref{eq:ab}).
The result is given by
\eq
\delta\mathscr{L}^{\Gamma}(x)+
\delta\mathscr{L}^{\Lambda\Lambda}(x)=
-\frac{N-2}{2}\,d(0)\dot e^\alpha(x_0)\dot e^\alpha(x_0)\, .
\en

\subsubsection*{ Figure~\ref{fig:NNLO}a, $\ell_1+\ell_2$}
Another two contributions in the diagram in Fig.~\ref{fig:NNLO}a come from the terms in the Lagrangian that contain the low-energy constants $\ell_1$ and $\ell_2$. These contributions are again given by the tadpoles:
\eq
\delta\mathscr{L}^{\ell_1}(x)+\delta\mathscr{L}^{\ell_2}(x)
=\frac{(4\ell_1+2\ell_2N)}{F^2}\,\ddot d(0)\dot e^\alpha(x_0)\dot e^\alpha(x_0)\, ,
\en
where
\eq
\ddot d(0)=-\int\frac{dk_0}{2\pi}\,\frac{1}{L^3}
\sum_{\bm{k}\neq 0}\frac{k_0^2}{k_0^2+\bm{k}^2}\, .
\en
Note that $\ddot d(0)=O(L^{-4})$ and hence the above contribution comes indeed
at NNLO.

  \subsubsection*{ Figure~\ref{fig:NNLO}c, $\Gamma R$}
The contribution of the two-loop the diagram shown in Fig.~\ref{fig:NNLO}c is given by
\eq
\delta\mathscr{L}^{\Gamma R}(x)=-\frac{1}{2F^2}\,\int d^4y\,\Gamma^{ii}(x_0)
\left(-d^2(x-y)\partial_\mu\partial_\mu d(0)+\partial_\mu d(x-y)\partial_\mu d(x-y)d(0)\right)\, .
\nonumber\\
\en
Taking into account that
\eq
\partial_\mu\partial_\mu d(0)&=&\int\frac{dk_0}{2\pi}\,\frac{1}{L^3}\,\sum_{\bm{k}\neq 0}
\frac{-k^2}{k^2}=0\, ,
\nonumber\\[2mm]
\int d^4y\,\partial_\mu d(x-y)\partial_\mu d(x-y)
&=&\int\frac{dk_0}{2\pi}\,\frac{1}{L^3}\,\sum_{\bm{k}\neq 0}
\frac{k^2}{(k^2)^2}=d(0)\, ,
\en
we finally get
\eq
\delta\mathscr{L}^{\Gamma R}(x)=-\frac{1}{2F^2}\,d^2(0)\left(
-(N-1)\dot e^\alpha(x_0)\dot e^\alpha(x_0)+\dot\Omega^{\alpha i}(x_0)\Omega^{\alpha i}(x_0)
\right)\, .
\en

  \subsubsection*{ Figure~\ref{fig:NNLO}d, $\Delta\Delta$}
The contribution of the the diagram shown in Fig.~\ref{fig:NNLO}d is given by
\eq
\delta\mathscr{L}^{\Delta\Delta}(x)=\frac{N-2}{4F^2}\,\int d^4y
\Delta^i(x_0)\Delta^i(y_0)d^2(x-y)\ddot d(x-y)\, .
\en
The integral above contains different scales, so the threshold expansion in $p_0$ is necessary:
\eq
&&\int d^3\bm{y} d^2(x-y)\ddot d(x-y)
\nonumber\\[2mm]
&=&\int d^3\bm{y}\int\frac{dk_{10}dk_{20}dk_{30}}{(2\pi)^3}\,
\frac{1}{L^9}\,\sum_{\bm{k}_1,\bm{k}_2,\bm{k}_3\neq 0}
\frac{-k_{30}^2e^{i(k_{10}+k_{20}+k_{30})(x_0-y_0)
  +i(\bm{k}_1+\bm{k}_2+\bm{k}_3)(\bm{x}-\bm{y})}}{k_1^2k_2^2k_3^2}
\nonumber\\[2mm]
&=&
\int\frac{dk_{10}dk_{20}dp_0}{(2\pi)^3}\,
\frac{1}{L^6}\,\sum_{\bm{k}_1,\bm{k}_2,\bm{k}_3\neq 0}
\frac{-\delta^3_{\bm{k}_1+\bm{k}_2+\bm{k}_3,0}(p_0-k_{10}-k_{20})^2e^{ip_0(x_0-y_0)}}{k_1^2k_2^2((p_0-k_{10}-k_{20})^2
  +(\bm{k}_1+\bm{k}_2)^2)}
\nonumber\\[2mm]
&=&\delta(x_0-y_0)\int\frac{dk_{10}dk_{20}}{(2\pi)^2}\,
\frac{1}{L^6}\,\sum_{\bm{k}_1,\bm{k}_2\neq 0}\frac{
  -\left(1-\delta^3_{\bm{k}_1+\bm{k}_2,0}\right)
(k_{10}+k_{20})^2}{k_1^2k_2^2((k_{10}+k_{20})^2
  +(\bm{k}_1+\bm{k}_2)^2)}+\cdots
\en
Carrying out the summation over the index $i$, we finally obtain
\eq\label{eq:diagram_d}
\delta\mathscr{L}^{\Delta\Delta}(x)=\frac{N-2}{F^2}\,e^\alpha(x_0)e^\alpha(x_0)
\int d^4y d^2(y)\ddot d(y)+\cdots
\en

  \subsubsection*{ Figure~\ref{fig:NNLO}e, $\Lambda R \Lambda$}
Finally, the contribution from the diagram Fig.~\ref{fig:NNLO}e can be written as
\eq
\delta\mathscr{L}^{\Lambda R \Lambda}(x)=\frac{2}{F^2}\,(\delta^{im}\delta^{jl}-\delta^{il}\delta^{jm})\int dy_0 \Lambda^{ij}(x_0)\Lambda^{lm}(y_0)K(x_0,y_0)\, ,
\en
where
\eq
K(x_0,y_0)&=&\int dz_0\int\frac{dp_0dq_0dk_0dl_0}{(2\pi)^4}\,
\frac{1}{L^6}\,\sum_{\bm{k}_1,\bm{k}_2,\bm{l}_1,\bm{l}_2\neq 0}
\delta^3_{\bm{k}_1+\bm{k}_2,0}\delta^3_{\bm{l}_1+\bm{l}_2,0}
\nonumber\\[2mm]
&\times&\frac{l_0l_\mu k_0k_\mu e^{ip_0(x_0-z_0)+iq_0(z_0-y_0)}}
           {((p_0-l_0)^2+\bm{l}_1^2)(l_0^2+\bm{l}_1^2)((q_0-k_0)^2+\bm{k}_1^2)(k_0^2+\bm{k}_1^2)}
           \nonumber\\[2mm]
     &=&\frac{1}{4}\,\delta(x_0-y_0)d^2(0)+\cdots
     \en
     Again, we have used threshold expansion in $p_0$ and $q_0$ to arrive at this result.
     Carrying out the summation over indices $i,j,l,m$, we finally get
     \eq
     \delta\mathscr{L}^{\Lambda R \Lambda}(x)=\frac{1}{2F^2}\,d^2(0)
     \left(\dot\Omega^{\alpha i}(x_0)\dot\Omega^{\alpha i}(x_0)-\dot e^\alpha(x_0)\dot e^\alpha(x_0)\right)\, .
     \en
As at the NLO, even if the set of the diagrams differs from the
ones considered in Ref.~\cite{Hasenfratz:2009mp}, one arrives exactly at the same
result
after adding all contributions and canceling terms with $\dot\Omega^{\alpha i}\dot\Omega^{\alpha i}$. Namely, the expression for the correction to
the moment of inertia at this order takes the form
\eq\label{eq:Hasenfratz1}
  \Theta=F^2L^3\left(1+\frac{C_1}{(FL)^2}+\frac{C_2}{(FL)^4}+\frac{C_3}{(FL)^4}\,
  \ln(FL)\right)\, ,
  \en
where $C_1,C_2,C_3$ are exactly the same as  in Ref.~\cite{Hasenfratz:2009mp}.
This result was of course expected, since  observables should not
depend on the choice of the interpolating field.\footnote{The logarithm
  in Eq.~(\ref{eq:Hasenfratz1}) emerges from the ultraviolet divergence in
the two-loop diagram in Fig.~\ref{fig:NNLO}d, see Eq.~(\ref{eq:diagram_d}). Details of calculations are given in Ref.~\cite{Hasenfratz:2009mp}.} Furthermore, it becomes clear that,
contrary to the claim of Ref.~\cite{Hasenfratz:2009mp}, the subleading terms in the
threshold expansion are only power-law suppressed and not exponentially suppressed.

The generalization of the approach to any order is crystal-clear. Evaluating
arbitrary Green functions, one first has to integrate out the fast modes corresponding to
the field $\bm{R}(x)$. Then, one obtains multi-scale Feynman integrals whose
counting in the small parameter $1/L$ is obscure.
The key observation is that the power counting can be formulated by
performing the threshold expansion in all integrals.
After that, calculations
can proceed without further ado. The vacuum matrix elements of the
product of the fields $\Omega^{\alpha\beta}(t)$ and derivatives thereof can be then
calculated along the lines described in Appendix~\ref{app:rotator}.
  Stated differently, the fast and slow modes are integrated out separately and the result
is ``glued together'' in order to arrive at the final expression for the Green functions.
Furthermore, as argued in Appendix~\ref{app:largescale}, the large momenta (of
order of $L^{-1}$) do not appear
in the matrix elements with the slow modes and thus threshold expansion is
not necessary there.

Finally, we make a short remark concerning  renormalization.
Potentially, ultraviolet divergences may arise at two different places: in the Feynman
diagrams with the fast modes, and in the infinite sums that are present in the Green
functions of the slow modes. Consider first the fast modes. The Feynman integrals
emerging here are identical to the ordinary ones, with the $\bm{k}=0$ component
removed. This removal, however, does not affect the ultraviolet divergences. On the other hand, as argued in Appendix~\ref{app:rotator}, all matrix elements containing slow modes are ultraviolet-finite. Hence, as expected, 
the couplings that are present in the effective
Lagrangian suffice to remove all ultraviolet divergences, and the pertinent $\beta$
functions coincide with the ones in the infinite volume.

\section{Conclusion and outlook}
\label{sec:concl}

Effective field theory methods allow one to study the temperature and volume dependence
of QCD, as well as different condensed-matter models, whose behavior in the
long-wavelength limit can be described by effective chiral Lagrangians. According to the
particular values of the parameters $T$ and $L$, as well as the lowest mass in the
system (the pion mass, $M$), the perturbative expansion of the physical observables in
these parameters should be rearranged, corresponding to what is termed as
different regimes.

The perturbative expansion of the effective theory in the $\delta$ regime, unlike
the $p$- and $\epsilon$-regimes, is characterized by the presence of two distinct
energy scales, corresponding to the so-called slow and fast pions. Such a separation is
unnecessary in the $p$-regime, whereas in the $\epsilon$-regime the slow mode is
not a dynamical variable. Only in the $\delta$-regime, in which the time-dependent
slow mode emerges, one is faced with the above-mentioned problem. For this
reason, the perturbative calculations, up to now, have been limited to the lower orders
in the expansion, and some cleverly designed tricks (like the field transformation considered in 
the present paper) were used to restore power counting at higher orders.

The key observation made in the present paper was that
the application of the threshold expansion to the Feynman integrals appearing in the perturbative expansion allows one to address the problem simultaneously at all orders and to rectify the power counting without resorting to further tricks. The known results for the rest-frame excitation spectrum at NNLO have been readily reproduced in a straightforward manner, demonstrating the independence of the physical observables on the choice of the interpolating field. The proposed method, on one hand, paves way for a systematic calculation of other observables in QCD and condensed matter physics even at higher orders and, on the other hand, can be generalized to the case, when the massive fermions interacting with pions, are present (this Lagrangian describes the physics of hole-doped antiferromagnets in the long-wavelength limit). In addition, an intriguing question arises, whether it is possible to implement the threshold expansion at the Lagrangian level using so-called
labeled fields (like in case of the heavy quark effective theory of QCD) that would render power counting explicit. These issues  form the subject of  future investigations.

\bigskip

{\em Acknowledgments:} The authors thank Thomas Becher, J\"urg Gasser, George Jackeli, Heinrich Leutwyler, Tom Luu, Johann Ostmeyer and Uwe Wiese for interesting discussions.
We would like to especially thank Martin Beneke for critical reading of the manuscript
and numerous suggestions that helped to improve the original draft considerably.
The work of was  funded in part by the Deutsche Forschungsgemeinschaft
(DFG, German Research Foundation) - Project-ID 196253076 - TRR 110.
AR acknowledges the financial support
from the Ministry of Culture and Science of North Rhine-Westphalia through the
NRW-FAIR project  and from the Chinese Academy of Sciences (CAS) President's
International Fellowship Initiative (PIFI) (grant no. 2024VMB0001).
The work of UGM was also supported  by the CAS President's International
Fellowship Initiative (PIFI) (Grant No.~2025PD00NN).

 \pagebreak
 
\appendix

\section{Wick's theorem for the slow modes?}
\label{app:Wick}

As already mentioned, Wick's theorem cannot be used if the free Lagrangian describes
the rigid rotator rather than a harmonic oscillator. We shall demonstrate this (rather obvious) statement here in an explicit example in the $O(3)$ model. Here, the eigenstates coincide with the eigenvectors of the angular momenta $|n\rangle=|\ell m\rangle$, and the eigenvalues are given by the expression $\varepsilon_\ell={\ell(\ell+1)}/{\Theta}$. To simplify
things as much as possible, we consider the Green functions of the fields $e^0(t)$. Then,
\eq
&&\langle\ell'm'|e^0|\ell m\rangle
=\int d\Omega Y^*_{\ell'm'}(\Omega)\sqrt{\frac{4\pi}{3}}Y_{10}(\Omega)Y_{\ell m}(\Omega)
\nonumber\\[2mm]
&=&\sqrt{(2\ell'+1)(2\ell+1)}(-1)^{m'}
\begin{pmatrix}
  1& \ell & \ell' \cr
  0 & m & -m'
\end{pmatrix}
\begin{pmatrix}
  1& \ell & \ell' \cr
  0 & 0 & 0
\end{pmatrix}\, .
\en
In the calculation of the Green functions of the field $e^0(t)$, we can set $m=m'=0$.
Using explicit values of the Wigner $3-j$ symbols, one gets
\eq
\langle 0,0|e^0|1,0\rangle&=&\langle 1,0|e^0|0,0\rangle=\frac{1}{\sqrt{3}}\, ,
\nonumber\\[2mm]
\langle 1,0|e^0|2,0\rangle&=&\langle 2,0|e^0|1,0\rangle=\frac{2}{\sqrt{15}}\, .
\en
With the use of the above formula, the two-point and the four-point functions of the
field $e^0(t)$ can be written as follows
\eq
\langle e^0(t_1)e^0(t_2)\rangle&=&\frac{1}{3}\,
\biggl(e^{-(\varepsilon_1-\varepsilon_0)(t_1-t_2)}\theta(t_1-t_2)
+e^{-(\varepsilon_1-\varepsilon_0)(t_2-t_1)}\theta(t_2-t_1)\biggr)
\nonumber\\[2mm]
&=&\frac{1}{3}\,\exp\left(-\frac{|t_1-t_2|}{\Theta}\right)\, ,
\en
and
\eq
&&\langle e^0(t_1)e^0(t_2)e^0(t_3)e^0(t_4)\rangle
=\sum_{{\sf perm}~ijkl}
\theta(t_i-t_j)\theta(t_j-t_k)\theta(t_k-t_l)
\nonumber\\[2mm]
&\times&\biggl[\frac{1}{9}\,\exp\biggl(-\frac{t_i-t_j+t_k-t_l}{\Theta}\biggr)
+\frac{16}{225}\,\exp\biggl(-\frac{t_i+2t_j-2t_k-t_l}{\Theta}\biggr)\biggr]\, ,
\quad\quad
\en
where ${{\sf perm}~ijkl}$ denotes all possible permutations of the indices.
It can be explicitly verified that the Wick's theorem does not hold. On the other hand,
considering the case of the harmonic oscillator, one may construct a similar four-point
function
\eq
&&\langle a(t_1)a(t_2)a^\dagger(t_3)a^\dagger(t_4)\rangle=
e^{-\omega(t_1+t_2-t_3-t_4)}
\nonumber\\[2mm]
&\times&\biggl[
2\sum_{ijkl=1234,2134,1243,2143}\theta(t_i-t_j)\theta(t_j-t_k)\theta(t_k-t_l)
\nonumber\\[2mm]
&+&\sum_{ijkl=1324,2314,1423,2413}\theta(t_i-t_j)\theta(t_j-t_k)\theta(t_k-t_l)\biggr]
\nonumber\\[2mm]
&=&e^{-\omega(t_1+t_2-t_3-t_4)}\biggl[\theta(t_1-t_3)\theta(t_2-t_4)+\theta(t_1-t_4)\theta(t_2-t_3)\biggr]\, .
\en
Here, $\omega$ denotes the  energy of a single excitation, and the matrix elements
of the creation/annihilation operators are normalized in the following way
\eq
\langle 0|a|1\rangle=\langle 1|a^\dagger|0\rangle=1\, ,\quad\quad
\langle 1|a|2\rangle=\langle 2|a^\dagger|1\rangle=\sqrt{2}\, .
\en
It is immediately seen that the Wick's theorem holds in the case of an oscillator.

To summarize, one sees that the validity of the Wick's theorem is tied to the choice of the
free Lagrangian. In our context, it is important to realize that, for this reason, it is
impossible to single out the contribution of slow modes at the level of individual
Feynman integrals, using, e.g.,
some kind of the threshold expansion, because slow modes are inherently non-perturbative.

\section{Field transformation}
\label{app:trafo}

Below, we consider the field transformation introduced in
 Refs.~\cite{Hasenfratz:1993vf,Hasenfratz:2009mp,Weingart:2010yv},
 in its continuum version.
 In order to specify the matrix $\Sigma(t)$ in Eq.~(\ref{eq:Sigma}), we first define the matrix $V(t',t)\in O(N)$ that
obeys the first-order differential equation
\eq\label{eq:diffeq}
\frac{\partial}{\partial t}\,V(t',t)=\Sigma(t')\Omega^T(t')
\frac{\partial}{\partial t}\,\Omega(t)\, .
\en
The following boundary condition is imposed
\eq
V(t,t)=\Sigma(t)\, ,
\en
where $\Sigma(t')$ is the matrix defined in Eq.~(\ref{eq:Sigma}), and
\eq\label{eq:hatSigma}
\hat\Sigma^{ij}(t)=V^{ij}(t',t)-\frac{V^{i0}(t',t)V^{0j}(t',t)}{1+V^{00}(t',t)}\, .
\en
These equations determine $\Sigma(t)$ any given time $t$, provided its initial value is
fixed. Note that the argument $t'$ in $\Sigma(t)$ is implicit (we remind the reader that
the boundary conditions are set at $t=t'$).

Let us now differentiate both sides of Eq.~(\ref{eq:hatSigma}) with respect to the variable
$t$ and consider the limit $t'\to t$ afterwards. The differentiation gives
\eq
&&\frac{\partial}{\partial t}\,\hat\Sigma^{ij}(t)\,=\,\frac{\partial}{\partial t}\,V^{ij}(t',t)
-\frac{V^{0j}(t',t)}{1+V^{00}(t',t)}\,\frac{\partial}{\partial t}\,V^{i0}(t',t)
\nonumber\\[2mm]
&-&\frac{V^{i0}(t',t)}{1+V^{00}(t',t)}\,\frac{\partial}{\partial t}\,V^{0j}(t',t)
+\frac{V^{i0}(t',t)V^{0j}(t',t)}{(1+V^{00}(t',t))^2}\,\frac{\partial}{\partial t}\,V^{00}(t',t)\, .
\en
On the other hand, from the boundary condition in Eq.~(\ref{eq:diffeq}) we get that
$V^{i0}(t,t)=V^{0j}(t,t)=0$ and $V^{00}(t,t)=1$. Hence,
\eq
\lim_{t'\to t}\frac{\partial}{\partial t}\,\hat\Sigma^{ij}(t)
=\lim_{t'\to t}\frac{\partial}{\partial t}\,V^{ij}(t',t)\, .
\en
Furthermore, the differential equation~(\ref{eq:diffeq}) at $t'\to t$ yields:
\eq
\lim_{t'\to t}\frac{\partial}{\partial t}\,\hat\Sigma^{ij}(t)
=\lim_{t'\to t}\hat\Sigma^{ik}(t)\left(\Omega^T(t')\frac{\partial}{\partial t}\,\Omega(t)\right)^{kj}\, ,
\en
or, finally,
\eq\label{eq:Sigma-Omega}
\left(\Sigma^T(t)\dot\Sigma(t)\right)^{ij}=\left(\Omega^T(t)\dot\Omega(t)\right)^{ij}\, ,
\en
where the limit $t'\to t$ on the left-hand side is implicit.

Next, let us consider the Lagrangian given in Eq.~(\ref{eq:L2}). After the field redefinition,
the quantity $\dot\Omega^{\alpha i}\Omega^{\alpha j}$ in the last term turns into
\eq
\left(
  \frac{d}{dt}\,
  (\Sigma\Omega^T)(\Omega\Sigma^T)
  \right)^{ij}
  &=&\left((\dot\Sigma\Omega^T+\Sigma\dot\Omega^T)\Omega\Sigma^T\right)^{ij}
\nonumber\\[2mm]
  &=&(\dot\Sigma\Sigma^T)^{ij}-\Sigma^{ik}(\Omega^T\dot\Omega)^{km}(\Sigma^T)^{mj}=0\, .
  \en
The last equality follows from Eq.~(\ref{eq:Sigma-Omega}).

Next, let us consider the term $\dot\Omega^{\alpha i}\dot\Omega^{\alpha j}$, which emerges in the same Lagrangian. Using field transformation and partial integration, it can be rewritten as
\eq
&&\frac{d}{dt}\,(\Omega\Sigma^T)^{\alpha i}
\frac{d}{dt}\,(\Omega\Sigma^T)^{\alpha j}
\nonumber\\[2mm]
&=&\frac{d}{dt}\,\biggl((\Omega\Sigma^T)^{\alpha i}
\frac{d}{dt}\,(\Omega\Sigma^T)^{\alpha j}\biggr)
-(\Omega\Sigma^T)^{\alpha i}
\frac{d^2}{dt^2}(\Omega\Sigma^T)^{\alpha j}\, .
\en
We have already shown that the first term vanishes. The second term can be rewritten as
\eq
&-&\lim_{t'\to t}(\Omega(t)\Sigma^T(t))^{\alpha i}
\frac{d^2}{d{t'}^2}\,(\Omega(t')\Sigma^T(t'))^{\alpha i}
=-\lim_{t'\to t}\frac{d^2}{d{t'}^2}\,(V^{ik}(t')(\Sigma^T(t'))^{kj})
\nonumber\\[2mm]
&=&-\lim_{t'\to t}\frac{d^2}{d{t'}^2}\,\biggl(
V^{ik}(t')\biggl(V^{jk}(t')-\frac{V^{j0}(t')V^{0k}(t')}{1+V^{00}(t')}
\biggr)\biggr)
\nonumber\\[2mm]
&=&-\lim_{t'\to t}\frac{d^2}{d{t'}^2}\,\frac{V^{i0}(t')V^{j0}(t')}{1+V^{00}(t')}\, .
\en
Taking now into account the boundary conditions on $V^{00},V^{i0},V^{0j}$ at $t'=t$, one
finally gets
\eq
-\lim_{t'\to t}(\Omega(t)\Sigma^T(t))^{\alpha i}
\frac{d^2}{d{t'}^2}\,(\Omega(t')\Sigma^T(t'))^{\alpha i}
=\frac{1}{2}\,\dot V^{i0}(t)\dot V^{j0}(t)\, .
\en
On the other hand,
\eq
\dot V^{i0}(t)=\lim_{t'\to t}\frac{d}{dt'}\left(\Sigma(t)\Omega^T(t)\Omega(t')\right)^{i0}
=(\Sigma(t)\Omega^T(t))^{i\alpha}\dot e^\alpha(t)\, .
\en
In particular, summing up over the indices $i,j$ we arrive at a simple result quoted in Refs.~\cite{Hasenfratz:1993vf,Hasenfratz:2009mp,Weingart:2010yv}:
\eq\label{eq:VV}
\dot V^{i0}(t)\dot V^{i0}(t)=\dot e^\alpha(t)\dot e^\alpha(t)\, ,
\en
from which Eq,~(\ref{eq:Hasenfratz}) directly follows.

\section{Calculation of the matrix elements with the slow modes}
\label{app:rotator}

As we have seen, the path integral over the slow modes cannot be calculated by using
the standard diagrammatic technique. The reason for this is that the unperturbed part of
the Lagrangian of the slow modes coincides with the Lagrangian of the rigid rotator rather
than that of an harmonic oscillator. As noted above, in order to evaluate the vacuum expectation
value of the operators containing slow modes, one may insert a full set of the eigenvectors
of the unperturbed Hamiltonian between each two operators. Consider, for example,
the Green function
\eq\label{eq:G}
G(t_1,\cdots,t_m)&=&\prod_t\int d^Ne(t)\delta(e^\alpha e^\alpha-1)
\exp\biggl(-\frac{\Theta}{2}\,\int dt\, \dot e^\alpha(t) \dot e^\alpha(t)\biggr)
\nonumber\\[2mm]
&\times&O_1\left[e(t_1)\right]\cdots O_m\left[e(t_m)\right]\, ,
\en
where $O_1,\ldots, O_m$ are arbitrary local operators built of $e^\alpha(t)$ (and time derivatives thereof which enter polynomially in the expression). Assume,
for instance that $t_1>t_2>\ldots>t_m$. Then, the above Green function can be represented by a sum
\eq
G(t_1,\cdots,t_m)&=&\sum_{n_1,\ldots,n_{m-1}}
e^{-\varepsilon_{n_1}(t_1-t_2)-\cdots-\varepsilon_{n_{m-1}}(t_{m-1}-t_m)}
\nonumber\\[2mm]
&\times&\langle 0|O_1|n_1\rangle\cdots\langle n_{m-1}|O_m|0\rangle\, .
\en
In case of the $O(N)$ rigid rotator, the eigenfunctions of the Laplace operator are given
by pertinent hyperspherical harmonics (see, e.g.,~\cite{Bateman-Erdelyi}). The polar
coordinates are introduced as
\eq
e^0&=&\cos\theta_1\, ,
\nonumber\\[2mm]
e^1&=&\sin\theta_1\cos\theta_2\, ,
\nonumber\\[2mm]
e^2&=&\sin\theta_1\sin\theta_2\cos\theta_3\, ,
\nonumber\\
&\cdots&
\nonumber\\
e^{N-2}&=&\sin\theta_1\cdots\sin\theta_{N-2}\cos\varphi\, ,
\nonumber\\[2mm]
e^{N-1}&=&\sin\theta_1\cdots\sin\theta_{N-2}\sin\varphi\, ,
\en
and the hyperspherical harmonics are given by
\eq\label{eq:Y}
Y(m_k;\theta_k,\pm\varphi)=N^{-1/2}(m_k)e^{\pm im_{N-2}\varphi}
\prod_{k=0}^{N-3}C_{m_k-m_{k+1}}^{m_{k+1}+\frac{N-1-k}{2}}
(\cos\theta_{k+1})\, .
\en
Here, the $m_k$ are the integers
\eq
n=m_0\geq m_1\geq\cdots m_{N-2}\geq 0\, ,
\en
where the index $n$ labels the irreducible representations (an analog of the angular
momentum $\ell$ in case of the $O(3)$ group).\footnote{The need for both signs $\pm$ in
Eq.~(\ref{eq:Y}) is related to our convention $m_{N-2}\geq 0$.} Furthermore, the $C_n^\mu(x)$ denote
Gegenbauer polynomials
\eq
C_n^\mu(x)=\frac{\Gamma(n+2\mu)}{\Gamma(n+1)\Gamma(2\mu)}
~_2F_1\left(-n,n+2\mu,\mu+\frac{1}{2},\frac{1}{2}-\frac{1}{2}\,x\right)\, .
\en
The normalization constant is given by
\eq
N(m_0,\cdots m_{N-2})&=&2\pi\prod_{k=1}^{N-2}E_k(m_{k-1},m_k)\, ,
\nonumber\\[2mm]
E_k(l,m)&=&\pi\frac{2^{k-2m-N+2}\Gamma(l+m+N-1-k)}
{\left(l+\frac{N-1-k}{2}\right)(l-m)!
  \left(\Gamma\left(m+\frac{N-1-k}{2}\right)\right)^2}\, .
\en
The eigenvalues $\varepsilon_n$ are given by Eq.~(\ref{eq:varepsilon_n}).

Assume first that the operator $O(e)$ does not contain time derivatives of $e^\alpha$.
Then, the matrix element of such an operator between two eigenstates characterized by
the sets $n\doteq\{m_k\}$ and $n'\doteq\{m_k'\}$ with $k=0,\ldots,N-2$ is given by
\eq
\langle n|O|n'\rangle
=\int d\Omega_NY^*(m_k;\theta_k,\pm\varphi)O\left[e(\theta_k,\varphi)\right]
Y(m_k';\theta_k,\pm\varphi)\, ,
\en
where $d\Omega_N$ denotes the volume in the $N$-dimensional space
\eq
d\Omega_N=(\sin\theta_1)^{N-2}\cdots \sin\theta_{N-2}d\theta_1\cdots d\theta_{d-2}d\varphi\, .
\en
In case when the operator $O\left[e\right]$ has a polynomial dependence on $e^\alpha$, the matrix elements $\langle n|O|n'\rangle$ are nonvanishing for selected values
on $n,n'$, owing to the Wigner-Eckart theorem. Note however that, for example,
the operator $\Omega^{ij}$ is not a polynomial in $e^\alpha$. On the contrary, the
dependence on time derivatives always has a polynomial form. Consider, for instance,
the operator $O=\tilde O\left[e\right]\dot e^\alpha$, where the operator
$\tilde O$ does not contain time derivatives. The matrix element of this operator
is given by
\eq
\langle n|O|n'\rangle=\sum_{n''}\langle n|\tilde O|n''\rangle(\varepsilon_{n'}-\varepsilon_{n''})\langle n''|e^\alpha|n'\rangle\, .
\en
The matrix element $\langle n''|e^\alpha|n'\rangle$ obeys selection rules in $n',n''$ and $\alpha$ (Wigner-Eckart theorem).\footnote{For example, in case of the $O(3)$ group, the
  eigenstates are labeled as $|n\rangle\doteq |\ell m\rangle$, and the matrix element
  $\langle\ell''m''|r^m|\ell'm'\rangle$ is non-zero, if and only if $\ell''=\ell'\pm 1$ and
  $m''=m+m'$. } Hence, the summation over $n''$ can be carried out explicitly.

To summarize, if all operators present in (\ref{eq:G}) are polynomials in $e^\alpha$,
all sums over intermediate states can be carried out in a closed form and contain only a finite number of of terms. From this it immediately follows that there are no
(ultraviolet) divergences in this case. Hence, the Green functions of $\Omega^{\alpha\beta}(t)$ with $\alpha=0$ and/or $\beta=0$ are ultraviolet-finite, as well as time derivatives
thereof to all orders. In case of a non-polynomial dependence, as in $\Omega^{ij}(t)$,
(infinite) sums remain in a final expression. These sums are anyway convergent
for $t_1\neq t_2\neq\cdots\neq t_m$, due to the presence of the exponential damping
factors. This argument does not apply, however, if any two (or more) arguments
coincide. Thus, the (potential) divergences in the position space must be proportional
to $\delta(t_i-t_j)$ and derivatives thereof and can  therefore be removed by local
counterterms in the effective Lagrangian that contain slow degrees of freedom only.
It is straightforward to see, however, that there is no need for such counterterms at
all. Indeed, consider first the case when no time derivatives are present. The matrix
elements of $\Omega^{ij}(t)$ itself are finite, since $e^ie^j/(1+e^0)$ is a regular
function even at $e^0\to -1$. Furthermore, at coinciding time arguments we merely get a product
of two (or more) operators $\Omega^{ij}(t)\Omega^{kn}(t)\ldots$, whose matrix
elements are also finite. Further, the derivative terms always contain
the quantity $\dot e^\alpha(t)$ polynomially and thus do not lead to the divergences as well. 
Finally, since the Lagrangian is built only of $\Omega^{\alpha\beta}$ and time derivatives thereof, 
we come to the conclusion that all Green functions of the fields
$\Omega^{\alpha\beta}(t)$ and their time derivatives are ultraviolet-finite.

\section{Is there a large scale present in the matrix elements with the slow modes?}
\label{app:largescale}

In this appendix, we shall argue that the threshold expansion carried out in the Feynman
integrals solves the problem with the violation of the power counting {\em everywhere,} 
i.e., there is no need for additional measures in the matrix elements containing slow modes. We shall explain the meaning of this statement in a particular example. No attempt will be made
to rigorously generalize it, albeit such a generalization seems to us to be
relatively straightforward.

\begin{figure}[t]
  \begin{center}
    \includegraphics*[width=9.cm]{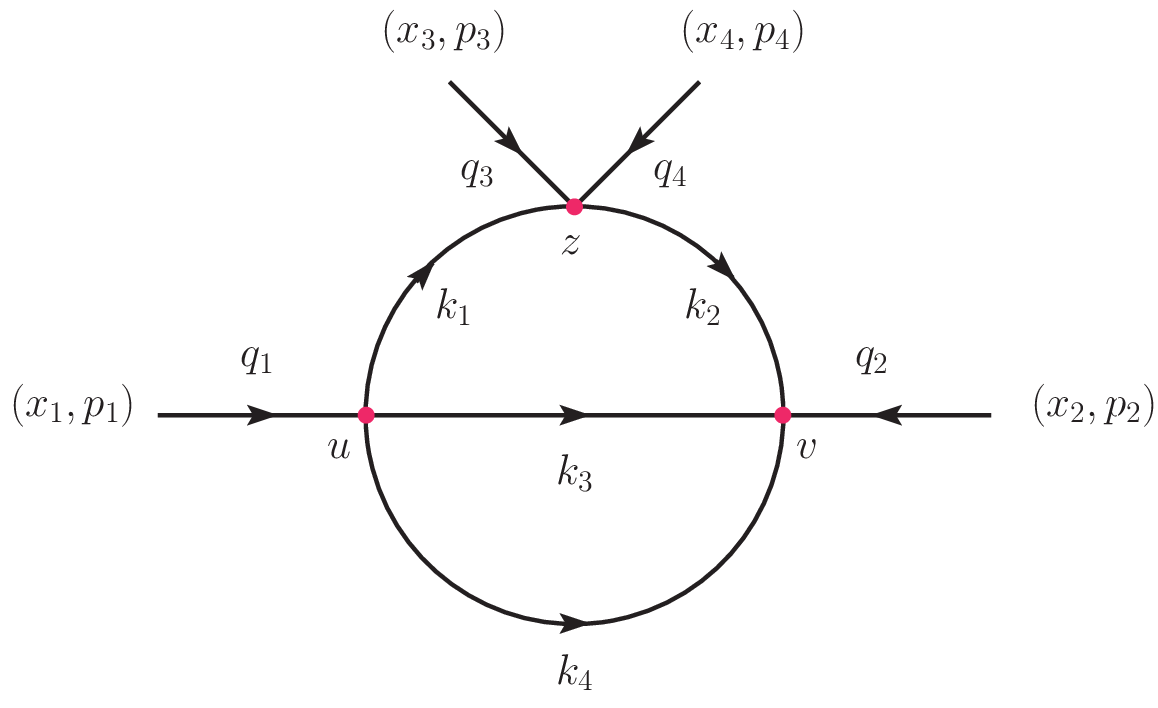}
    \caption{A typical two-loop diagram. Notations for momenta and vertices
      are the same as in the text.}
    \label{fig:typical}
  \end{center}
  \end{figure}

For definiteness, let us consider the following four-point function
\eq
G_4(\{p_i\})=\int \prod_i^4 d^4x_ie^{-ip_ix_i}
\langle S(x_1)S(x_2)S(x_4)S(x_4)\rangle\, .
\en
Note that, in order to ease the notations, we shall discard all indices, derivative couplings, overall normalization factors, etc. For example, the operator $S(x)$ is given by
a product $S(x)=\Omega(x_0)R(x)$. We shall further concentrate on a typical two-loop contribution to this Green function shown in Fig.~\ref{fig:typical}. The interaction Lagrangian in the vertices
will be chosen in a simple form $\mathscr{L}_{\sf int}(x)=R^4(x)O(x_0)$, where the operator $O$ collects soft modes. Integrating out fast modes leads to the following expression:
\eq\label{eq:convolution}
G_4(\{p_i\})&=&
\int \prod_i^4 dx_{i0}e^{-ip_{i0}x_{i0}}
\int du_0 dv_0 dz_0 \langle\Omega(x_{10})\cdots \Omega(x_{40})
O(u_0)O(v_0)O(z_0)\rangle
\nonumber\\[2mm]
&\times& K(\{\bm{p}_i\},\{ x_{i0}\};u_0,v_0,z_0)\, ,
\en
where
\eq
K(\{\bm{p}_i\},\{ x_{i0}\};u_0,v_0,z_0)&=&\int d^3\bm{x}_1\cdots d^3\bm{x}_4e^{-i(\bm{p}_1\bm{x}_1+\cdots+\bm{p}_4\bm{x}_4)}
\nonumber\\[2mm]
&\times&d(x_1-u)d^2(u-v)d(x_2-v)d(x_3-z)d(x_4-z)\, .
\en
Next, we substitute the Fourier-transform for each propagator of the fast
mode~(\ref{eq:propagator-i}). After this substitution, the
argument of the exponential $e^{iA}$ containing only zeroth components of
momenta is given by\footnote{The momenta are defined in Fig.~\ref{fig:typical}.}
\eq
A&=&q_{10}x_{10}+q_{20}x_{20}+q_{30}x_{30}+q_{40}x_{40}
+(-q_{10}+k_{10}+k_{30}+k_{40})u_0
\nonumber\\[2mm]
&+&(-q_{20}-k_{20}-k_{30}-k_{40})v_0
+(-q_{30}-q_{40}-k_{10}+k_{20})z_0\, .
\en
Next, we define the ``small'' momenta $p_0,q_0$, corresponding to the slow mode:
\eq
p_0&=& k_{10}+k_{30}+k_{40}-q_{10}\, ,
\nonumber\\[2mm]
q_0&=& -k_{20}-k_{30}-k_{40}-q_{20}\, ,
\nonumber\\[2mm]
\Delta&=&
q_{30}+q_{40}+k_{10}-k_{20}-p_0-q_0
=q_{10}+q_{20}+q_{30}+q_{40}\, .
\en
The Feynman integral, corresponding to the two-loop diagram in Fiq.~\ref{fig:typical},
is given by
\eq
I&=&\int \frac{dk_{30}dk_{40}}{(2\pi)^2}\,\frac{1}{L^6}\,\sum_{\bm{k}_3,\bm{k}_4\neq 0}
\frac{1}{((p_0+q_{10}-k_{30}-k_{40})^2+(\bm{q}_1-\bm{k}_3-\bm{k}_4)^2)}
\nonumber\\[2mm]
&\times&\frac{1}{((q_0+q_{20}+k_{30}+k_{40})^2+(\bm{q}_2+\bm{k}_3+\bm{k}_4)^2)
  (k_{30}^2+\bm{k}_3^2)  (k_{40}^2+\bm{k}_4^2)}\, .
\en
Applying threshold expansion amounts to expanding the integrand in powers of
$p_0,q_0$. At the first order, $I$ does not depend on these variables at all, and a
subsequent integration of
the exponent $e^{iA}$ over $p_0,q_0$ yields $\delta$-functions:
\eq
E=\exp\left(i\sum_{i=1}^4q_{i0}x_{i0}-iz_0\Delta\right)
\delta(u_0-z_0)\delta(v_0-z_0)+\cdots
\en
In higher orders, the derivatives of the delta functions will emerge.

At the next step, the result should be convoluted with the matrix element of the slow
operators in Eq.~(\ref{eq:convolution}). Potential danger arises from the factor
$e^{-iz_0\Delta}$, because $\Delta$ contains the energies of the fast modes and the
convolution could inject the large momentum into the matrix element of the slow
modes. It is easy to see, however, that this does not happen. Indeed, using
translational invariance of the matrix element, one can shift all arguments of the
operators by $z_0$. The dependence on $z_0$ in the factor $E$ disappears and
instead emerges in the exponential
$\exp\left(i\sum\limits_{i=1}^4p_{i0}x_{i0}\right)$, which is also present
Eq.~(\ref{eq:convolution}). Carrying out integration over $z_0$ yields a trivial delta-function,
corresponding to the conservation of total energy, with no hard scales present in the
remainder.

As mentioned above, we make no attempt here to apply the same argument to a generic
multi-loop diagram, albeit the fact that the argument is based only on the   
conservation of energy makes it very likely that this may work in other cases as well.
Putting differently, it is known that the large momenta should be conserved separately.
Here, we claim that these large momenta can be routed through the fast lines so that
they are never injected in the matrix elements of the slow modes.
A detailed investigation of this claim, however, forms a subject of a
separate investigation, and we plan to undertake it in the future.

\bibliographystyle{JHEP}
\bibliography{ref}

\end{document}